\documentclass[a4paper,12pt]{article}
\pdfoutput=1
\usepackage[utf8]{inputenc}
\usepackage{amsmath}
\usepackage{amsfonts}
\usepackage{amssymb}
\usepackage{graphicx, rotating}
\usepackage{latexsym}
\usepackage{color}
\usepackage{bm}
\usepackage{subcaption}
\usepackage[colorlinks,allcolors=blue]{hyperref}
\usepackage[numbers,sort&compress]{natbib}

\usepackage[english]{babel}

\newcommand{\Slash}[1]{{\ooalign{\hfil#1\hfil\crcr\raise.167ex\hbox{/}}}}

\newcommand{\beq}{\begin{equation}}  \newcommand{\eeq}{\end{equation}}
\newcommand{\bef}{\begin{figure}}  \newcommand{\eef}{\end{figure}}
\newcommand{\bec}{\begin{center}}  \newcommand{\eec}{\end{center}}
  
\newcommand{\laq}[1]{\label{eq:#1}}  

\newcommand{\Eq}[1]{Eq.~(\ref{eq:#1})}
\newcommand{\Eqs}[1]{Eqs.~(\ref{eq:#1})}

\newcommand{\Sec}[1]{Sec.\ref{chap:#1}}

\newcommand{\vev}[1]{ \left\langle {#1} \right\rangle }

\newcommand{\lac}[1]{\label{chap:#1}}
\newcommand{\SU}[1]{{\rm SU{#1} } }

\newcommand{\pqty}[1]{\left( #1 \right)}

\newcommand{\rqty}[1]{\left[ #1 \right]}
\newcommand{\absqty}[1]{\left| #1 \right|}
\def\({\left(}
\def\){\right)}

\def\O{\mathcal{O}}
\def\U{\mathop{\rm U}}

\newcommand{\AND}{~{\rm and}~}

\newcommand{\MEV}{ {\rm \, MeV} }
\newcommand{\GEV}{ {\rm \, GeV} }
\newcommand{\TEV}{ {\rm \, TeV} }

\def\a{\alpha}

\def\d{\delta}
\def\e{\epsilon}
\def\f{\phi}
\def\g{\gamma}
\def\h{\theta}

\def\m{\mu}
\def\n{\nu}

\def\s{\sigma}
\def\t{\tau}

\def\x{\xi}
\def\y{\eta}

\def\D{\Delta}
\def\G{\Gamma}
\def\L{\Lambda}
\def\F{\Phi}

\def\ol{\overline}
\def\tl{\tilde}
\def\*{\dagger}

\def\pmass{M_{\rm pl}}

\begin{document}
\renewcommand\bibname{\Large References}

\begin{flushright}
{TU-1260}
\end{flushright}

\begin{center}

\vspace{0.1cm}

{\Large\bf {Superheavy dark matter from the natural inflation in light of the highest-energy astroparticle events}}
\vspace{0.5cm}

{\bf Kohta Murase}$^{1,2}$, {\bf Yuma Narita}$^{3,4}$ and {\bf Wen Yin}$^{3,4}$\\
 \vspace{0.5cm}
{\em  (1)Department of Physics; Department of Astronomy \& Astrophysics; Center for Multimessenger Astrophysics, Institute for Gravitation and the Cosmos, The Pennsylvania State University, University Park, PA 16802, USA}\\
{\em (2) Center for Gravitational Physics and Quantum Information, Yukawa Institute for Theoretical Physics, Kyoto University, Kyoto, Kyoto
606-8502, Japan}\\
{\em 
(3) Department of Physics, Tokyo Metropolitan University,
 Hachioji-shi, Tokyo 192-0397 Japan} \\
 {\em 
(4) Department of Physics, Tohoku University, Sendai, Miyagi 980-8578, Japan}

\abstract{
Superheavy dark matter has been attractive as a candidate of particle dark matter. We propose a ``natural" particle model, in which the dark matter serves as the inflaton in natural inflation, while decaying to high-energy particles at energies of $10^{9}-10^{13} \, \text{GeV}$ from the prediction of the inflation. A scalar field responsible for diluting the dark matter abundance revives the natural inflation either with or without the recent data by the Atacama Cosmology Telescope (ACT) and baryon acoustic oscillation results from Dark Energy Spectroscopic Instrument.
Since the dark matter must be a spin-zero scalar, we carefully study the galactic dark matter 3-body decay into fermions and two body decays into a gluon pair, and point out relevant multi-messenger bounds that constrain these decay modes. Interestingly, the predicted energy scale may coincide with the AMATERASU event and/or the KM3NeT neutrino event, KM3-230213A. We also point out particle models with dark baryon to further alleviate $\gamma$-ray bounds. This scenario yields several testable predictions for the UHECR observations, including the highest-energy neutrons that are unaffected by magnetic fields, the tensor-to-scalar ratio, the running of spectral indices, $\alpha_s\gtrsim\mathcal{O}(0.001)$, and the existence of light new colored particles that could be accessible at future collider experiments.
Further measurements of high energy cosmic rays, including their components and detailed coordinates may provide insight into not only the origin of the cosmic rays but also inflation. 
}

\end{center}
\clearpage
\section{Introduction}
Particle properties of dark matter (DM) are completely unknown despite recent significant advancements. The mass of DM can be much heavier than the TeV or even PeV scale especially for decaying DM whose mass can go beyond the unitarity bound. Recently, such superheavy DM~\cite{Greene:1997ge,Chung:1998zb,Chung:1998ua,Chung:1999ve,Chung:2001cb,Chung:1998bt,Kolb:2007vd,Kannike:2016jfs,Ling:2021zlj} has been of much interest in light of high-energy multi-messenger observations relying on ultrahigh-energy cosmic rays (UHECRs), including ultrahigh-energy (UHE) photons, as well as UHE neutrinos (e.g., Refs.~\cite{Ellis:1990nb,Gondolo:1992cw,Berezinsky:1997hy,Birkel:1998nx,Kuzmin:1998uv,Kuzmin:1999zk,Kachelriess:2007aj,Yuksel:2007ac, Murase:2012xs,Esmaili:2012us} and Refs.~\cite{Feldstein:2013kka,Esmaili:2013gha,Murase:2015gea,Aloisio:2015lva,Cohen:2016uyg, Hiroshima:2017hmy,Kachelriess:2018rty, Ishiwata:2019aet,Jaeckel:2020oet,Guepin:2021ljb,Arguelles:2022nbl,Das:2023wtk,Fiorillo:2023clw,Das:2024bed} after the discovery of IceCube neutrinos).

The origin of UHECRs with $\gtrsim10^{18.5}\,\text{eV}$ is one of the most enduring and unresolved mysteries in modern astrophysics, a challenge that has driven decades of research into the extreme acceleration mechanisms and source environments responsible for their production. Conventional astrophysical candidates for UHECR sources include active galactic nuclei, $\gamma$-ray bursts, and other powerful accelerators capable of satisfying the Hillas criterion~\cite{Hillas:1985is,Anchordoqui:2018qom,Kotera:2011cp}. 
A crucial theoretical consideration for such extremely high-energy events is the Greisen-Zatsepin-Kuzmin (GZK) cutoff at $(4$--$5)\times10^{19}\,\text{eV}$~\cite{Greisen:1966jv,Zatsepin:1966jv}, which arises from interactions between cosmic rays and the cosmic microwave background (CMB). Not only UHECR protons but also UHECR nuclei lose energy rapidly via the photodisintegration as they propagate over cosmological distances. 
Consequently, the highest-energy cosmic rays must originate from sources within $\mathcal{O}(10-100)\,\text{Mpc}$ of Earth to avoid significant energy degradation, and the highest-energy cosmic rays with energies above $10^{20}\,\text{eV}$ have been studied extensively by experiments such as the Telescope Array (TA)~\cite{TelescopeArray:2018xyi,TelescopeArray:2023sbd} and the Pierre Auger Observatory (Auger)~\cite{PierreAuger:2020kuy}.  
Despite this distance limitation, some events are detected from directions lacking any obvious active source, such as the AMATERASU event by TA~\cite{TelescopeArray:2018xyi,TelescopeArray:2023sbd}. 

High-energy neutrino astrophysics is now realized thanks to observations by the IceCube neutrino experiment at the south pole. Recently, an interesting high-energy neutrino event was reported by the KM3NeT Collaboration~\cite{KM3NeT:2025npi}. This event, KM3-230213A, has an estimated energy of approximately $220$~PeV, making it the highest-energy neutrino event candidate recorded to date. Its origin is still under debate, and if this is not the background, the inferred flux level is in strong tension with the IceCube data~\cite{IceCube:2018fhm,IceCube:2020wum,IceCube:2025ezc} and Auger data~\cite{PierreAuger:2023pjg} (see Ref.~\cite{Li:2025tqf} for details).  
Although the IceCube-KM3Net joint flux is totally consistent with many predictions of astrophysical models~\cite{Kotera:2025jca}, astrophysical scenarios have difficulty in explaining the KM3Net data point without the IceCube upper limit, interpretations relying on physics beyond the Standard Model (BSM) have also been discussed~\cite{Borah:2025igh,Kohri:2025bsn,Narita:2025udw,Jiang:2025blz,Jho:2025gaf,Barman:2025hoz,Klipfel:2025jql}. 

The observations of the AMATERASU and KM3-230213A events give new motivations for exploring superheavy DM using UHECRs and UHE neutrinos. Particle physics of superheavy DM should address these key questions:
\begin{enumerate}
    \item Why would the DM possess such an enormous mass scale?
    \item How does the DM acquire the correct abundance?
    \item What are the underlying BSM models?
\end{enumerate}
Motivated by the numerical coincidence between the inflaton mass scale in high-scale inflation and the required DM mass scale for the observed UHE events, we propose inflaton DM models to answer the above questions (1-3), in which rare decay into SM particles or dark particles. 

Cosmic inflation~\cite{Starobinsky:1979ty,Starobinsky:1980te,Guth:1980zm,Sato:1980yn,Linde:1981mu,Mukhanov:1981xt,Hawking:1981fz,Chibisov:1982nx,Hawking:1982cz,Guth:1982ec,Albrecht:1982wi,Starobinsky:1982ee}, which generates the primordial density perturbations, is strongly supported by recent CMB observations~\cite{Planck:2018jri,Planck:2018vyg}. From a field-theory perspective, the inflation requires a scalar field (the inflaton) with a very flat potential to drive slow-roll inflation. 

The scenario of the inflaton DM in the context of multi-natural inflation~\cite{Czerny:2014wza, Czerny:2014xja,Czerny:2014qqa,Higaki:2014sja,Narita:2023naj} has been considered, where an axion-like particle (ALP), i.e., a pseudo Nambu-Goldstone boson, has a potential with two or more cosine terms for driving the inflation. Such ALPs (that we often call axions) have been considerred as nice candidates for both inflation and DM, but they are usually assumed to be different particles. The mass scale of the axion is generated by non-perturbative effect and has a logarithmically scaling.
Surprisingly, the minimal scenario has an interesting parameter region, where a single axion explains both inflation and DM, which was called as the ALP miracle scenario~\cite{Daido:2017wwb,Daido:2017tbr,Takahashi:2023vhv}. 
In that scenario, one assumes the axion potential has the upside-down symmetry, so that the potential hilltop, where the inflation takes place, has the same curvature as the bottom, which provides with the axion mass. It was found that the reheating is successful if the axion mass, and also the inflationary Hubble parameter, which is around the order of the hilltop curvature, is eV or so. Because of their small mass scales, the axion can be metastable, and thus may become DM in the present Universe. Reheating is successful due to the time-dependent effective mass and the dissipation effect, and some of the inflaton may remain as DM. 
However, simple natural inflation models, where the axion potential has a single cosine term,~\cite{Freese:1990rb,Adams:1992bn}, are not considered for inflaton DM models.  
The first reason is that the original natural inflation model has already been disfavored by current cosmic microwave background (CMB) data due to its prediction of a too-large tensor-to-scalar ratio and a super-Planckian decay constant, which is questioned in the quantum gravity view point. 
The second reason is that the inflaton is too heavy to be long-lived enough to be the DM unless there is a symmetry. 

In this paper, we point out that the natural inflation potential satisfying an approximate dark charge conjugation symmetry can stabilize the inflaton as DM. Here we show a new scenario, in which the natural inflation model revives, by introducing a temporally ``dark energy", which reheats the universe later and dilute the inflaton to be the dominant DM. As in the ALP miracle scenario, the upside-down symmetric potential requires the inflationary Hubble parameter to be close to the vacuum mass parameter. It turns out that the natural parameter region has the DM mass in the range of $10^9$-$10^{13}\,$GeV.
We then study the decay of the inflaton DM into Standard Model (SM) particles. Because the inflaton is a heavy scalar, its two-body decay into a fermion pair is chirality suppressed, making three-body decays or decays into a gauge boson pair particularly relevant. We perform a detailed analysis of the multi-messenger constraints, including $\gamma$-rays, (anti)nucleons, and (anti)neutrinos, arising from these decay channels. 
We find that within the allowed mass range the scenario can explain the AMATERASU within the $2\sigma$ range. 
With a different mass of inflaton within the parameter region, the combined flux from the KM3-230213A event and the IceCube upper limit could be explained within the 1$\sigma$ level by the 3-body leptonic decay. 
 
Since explainning the AMATERASU or other high-energy astroparticle events often violates $\gamma$-ray observations, we propose an extension of the model in which the inflaton predominantly decays into dark-sector particles, which cascades to dark baryons which decay into nucleons with other SM particles suppressed due to a symmetry and mass degeneracy (see also the alleviation of the $\gamma$-ray bound for the IceCube data \cite{Hiroshima:2017hmy} and KM3-230213A event \cite{Narita:2025udw}). In this extended scenario, the AMATERASU event as well as the significant components of the highest energy neuclei events could be attributed to the DM decay. 

Our model yields several testable predictions, including a lower bound on the tensor-to-scalar ratio, running of the spectral index, UHE (anti)nuetrons, and the existence of light new colored particles in the dark baryon models that may be observable at future collider experiments.

This paper is organized as follows. 
In \Sec{1}, we examine the inflaton-DM model in detail and demonstrate its consistency with both CMB data on inflation and the DM abundance. \Sec{2} discusses the coupling of the inflaton to SM particles and presents the constraints on decay channels induced by the small explicit breaking of the dark charge conjugation symmetry. 
We also perform a study in the context of KM3-230213A event. 
In \Sec{PM}, we construct a model in which DM decays into dark-sector particles. The last section is devoted to conclusions and discussion.

\section{Energy scales of astroparticles from natural inflaton}
\lac{1}
Let us describe our model. First, we consider the model of natural inflation driven by a field, $\phi$. Next, we discuss a mechanism, by which $\phi$ is produced as DM.

\subsection{Inflationary dynamics}
The natural inflation has been constrained by different observations. One of the simplest possibilities to save the natural inflation is to introduce a constant term. For example, the inflaton potential is given by 
\beq
\laq{save}
V(\f)= \L^4 \(\gamma + 1 - \cos\(\frac{\f}{f_\f}\) \),
\eeq
where $\Lambda$ is the potential energy and $f_\phi$ is the decay constant of $\phi$. The key difference from the ordinary natural inflation is the existence of the parameter $\gamma$ that is positive.
This term ensures that the vacuum energy remains non-zero during the inflation, which means that it does not correspond to the true vacuum in which we live. Therefore, we should consider models, in which the positive vacuum energy eventually vanishes after the inflation. An example of such a model is hybrid natural inflation~\cite{Ross:2009hg,Ross:2010fg,Ross:2016hyb} (see also \cite{Narita:2023naj}), which introduces an additional scalar field, $\Psi$. 
Alternatively, $\Psi$ may drive another inflation, e.g., Refs.~\cite{Bedroya:2019snp,Berera:2019zdd,Sasaki:2018dmp}.
This field has a role of not only enhancing the vacuum energy during inflation but also reheating the universe after the inflation. Assuming that $\Psi$ is also responsible for ending the natural inflation, it is sufficient to discuss the dynamics of the single field $\phi$ to derive the observables of CMB data. We will study this part that does not depend on $\Psi$.

In our model, the slow-roll parameters are given by
\begin{align}
\label{eq:slowrollpara1}
\e &\equiv \frac{\pmass^2}{2} \pqty{\frac{V'(\f)}{V(\f)}}^2 = \frac{\pmass^2}{2 f_\phi^2} \frac{\sin^2 \theta}{(\gamma + 1 - \cos \theta)^2}, \\
\label{eq:slowrollpara2}
\y &\equiv \pmass^2 \frac{V''(\f)}{V(\f)} = \frac{\pmass^2}{f_\phi^2} \frac{\cos{\theta}}{\g + 1 - \cos \theta},
\\
\label{eq:slowrollpara3}
\x^2 &\equiv \pmass^4 \frac{V'(\f)}{V(\f)} \frac{V'''(\f)}{V(\f)} = - \frac{\pmass^4}{f_\phi^4} \frac{\sin^2 \h}{(\gamma + 1 - \cos \theta)^2},
\end{align}
where $\h = {\f}/{f_\f}$ is a dimensionless parameter of the field $\phi$. Here, a prime symbol denotes the ordinary derivative with respect to $\f$.
For $\g = \mathcal{O}(1)$, this scenario does not work due to the constraint of the running of spectral index or the tensor-to-scalar ratio as discussed below.
Therefore, we assume $ \gamma \gg 1 $ for the successful inflation.
Then, we get the following in simple form:
\begin{equation}
    \e \simeq \frac{\sin^2 \h}{2 c \g}, \,
    \y \simeq \frac{\cos{\h}}{c}, \,
    \x^2 \simeq - \frac{\sin^2 \h}{c^2},
\end{equation}
where we defined the parameter $c$ as
\begin{equation}
\label{eq:c}
c \equiv \g \frac{f_\f^2}{\pmass^2}.
\end{equation}
 We need these parameters to estimate the CMB observations, such as the spectral index $n_s$, the tensor-to-scalar ratio $r$, and the running of spectral index $\a_s$. For single-field slow-roll inflation, the formulae of these indices are
\begin{align}
\label{eq:spectralindex}
n_s &= 1+ 2 \y_* - 6 \e_* , \\
\label{eq:tensortoscalar}
r &= 16 \e_*, \\
\label{eq:running}
\a_s &= 16 \e_* \y_* -24 \e_*^2 - 2\x^2_*.
\end{align}
Here, we use the symbol ``$*$" to denote the quantity at the horizon exit of the CMB scale. For $\g \gg 1$, it is clear that $\e$ is much smaller than $|\y|$ and $\sqrt{|\x^2|}$ for any value of $\h$ from \Eqs{slowrollpara1}-\eqref{eq:slowrollpara3}, leading to simple expressions for the spectral index and the running of spectral index: $n_s \simeq 1 + 2 \y_*$ and $\a_s \simeq -2 \x^2_*$. Each CMB observation can constrain the parameters in the potential \eqref{eq:save}. So, we will examine a parameter region with respect to $c$ and $f_\f$ to be consistent with the CMB observations. 

\begin{figure}[t!]
    \begin{center}
      \includegraphics[width = 81mm]{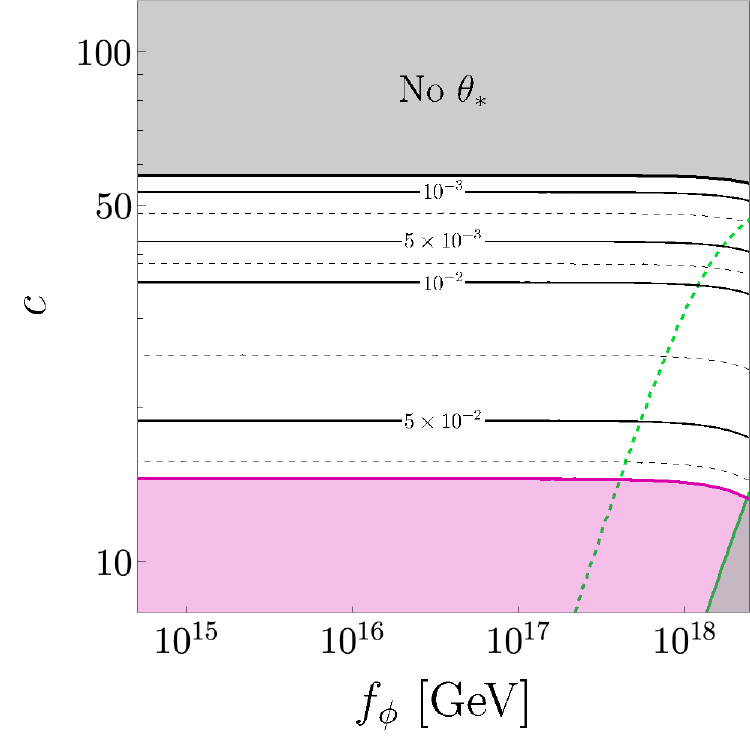}  
    \end{center}
  \caption{The contour of the running $\a_s$ in the viable parameter space.
  In the gray shaded region in the top, this scenario does not achieve a field value $\theta_*$ consistent with the observed spectral index. This figure also shows constraints from the tensor-to-scalar ratio (green shaded region in the right bottom) and the running of the spectral index (red shaded region in the bottom) required for successful inflation. The green dashed line represents the contour for $r=0.001$, up to which could be probed by future measurements.}
  \label{fig:conts}
\end{figure}

Fig.\ref{fig:conts} shows the parameter region in $f_\f$-$c$ plane, along with the predictions for the tensor-to-scalar ratio and the running of the spectral index. The contours as well as the colored excluded regions will be discussed in detail in the following. 
$f_\phi\gtrsim 10^{15}\GEV$, which we show in the figure, is favored from the context of hierarchy problem between the $f_\f$ and the Planck scales. Either if $\f$ is from extradimensional theory, such as string theory, or from a spontaneous symmetry breaking in quantum field theory without introducing hierarchy problem between the breaking scale and the Planck scale, this range is preferred.\footnote{
Although $\g$ is much larger than 1, our model does not have other fine-tuning for having a low scale inflation since the cosine potential is dynamically generated (note also that various relevant formulas are in the form of $c\sim 10$ for any scales).} 
Thus, the figure shows the natural region for the scenario. 

The parameter region is obtained from the following constraints.
First, we determine the value of $\h_*$ which can account for the observed spectral index \cite{Planck:2018jri}: 
\begin{equation}
\label{eq:nsobs}
n_{s,\rm obs} = 0.965 \pm 0.004,
\end{equation}
where we have adopted the result of the {\it Planck} TT, TE, EE + lowE. 
Based on the recent data from the ACT collaboration~\cite{ACT:2025fju, ACT:2025tim}, combined with CMB measurements from BICEP/Keck~\cite{BICEP:2021xfz} and Planck~\cite{Planck:2018jri, Planck:2018vyg}, as well as baryon acoustic oscillation results from DESI~\cite{DESI:2024mwx}, the spectral index have been measured to be $n^{\rm ACT}_{s,\text{obs}} = 0.974 \pm 0.003$. Even if this value is adopted, the following discussion remains valid.
For $\g \gg 1$, \Eq{spectralindex} suggests that $\theta_*$ is governed by the equation
\begin{equation}
\label{eq:thetacmb}
\cos{\h_*} \simeq \frac{c}{2} (n_{s,\rm obs} - 1).
\end{equation}
Therefore, we can get the field value at the CMB horizon exit from this equation. However, for the sufficiently large value of $c$, there may exists no solution for $\theta_*$ in Eq.~\eqref{eq:thetacmb}. The condition under which this scenario could have a field value consistant with the observed spectral index provides an upper limit of $c$:
\begin{equation}
\label{eq:nscondition}
c < \frac{2}{1 - n_{s,\rm obs}} \sim 57.
\end{equation}
Here, at the last equality, $n_{s,\rm obs}$ is substituted with the central value of the observed spectral index from \Eq{nsobs}. {If $c$ is larger than $57$, the potential cannot induce the red-tilted power spectram as implied by CMB data. This is changed to $\sim 77$ by using $n_{\rm s,obs}^{\rm ACT}$, and we have even larger parameter region. 

The gray region in Fig.~\ref{fig:conts} represents the parameter space where $\theta_*$ does not exist in the potential.}

We need to be careful about the running $\a_s$ for the relatively small value of $c$. Since only $\e$ is proportional to the inverse of $\g$, the terms involving $\e$ in Eq.~\eqref{eq:running} can be negligible. So, the running $\a_s$ can be approximately written as 
\begin{equation}
\label{eq:asapp}
\a_s \simeq - 2 \x^2_* \simeq 2 \frac{\sin^2{\h_*}}{c^2}.
\end{equation}
It tells us that this scenario predicts the positive running. Since $\theta_*$ is expected from the observed spectral index, we get
\begin{equation}
\label{eq:aspredict}
\a_s \simeq \frac{2}{c^2} - \frac{(n_{s ,\rm obs}-1)^2}{2}.
\end{equation}
The CMB constraint on this index is 
\begin{equation}
\label{eq:ascmb}
\a_{s, \text{obs}} = - 0.0045 \pm 0.0067, 
\end{equation}
where we have adopted the result of the {\it Planck} TT, TE, EE + lowE \cite{Planck:2018jri}. Here, we allow the running up to the $2 \sigma$ uncertainty, enabling us to set its upper limit in this analysis. It provides the strong restriction of the parameter $c$,
\begin{equation}
\label{eq:ascondition}
c \gtrsim 14.
\end{equation}
On the other hand, the ACT result gives $\a_{s, \text{obs}}^{\rm ACT}= 0.0062 \pm 0.0052$ which leads to $c\gtrsim 11.$ The new data slightly favors the positive running which is the prediction of our scenario, because satisfying $c<2/(1-n_{\rm s,obs}),$ $\alpha_s$ is positive.
Note that, future measurements of the curvature power spectrum can test our scenario through the running, e.g., in CMB-S4~\cite{CMB-S4:2016ple}, SPHEREx~\cite{Dore:2014cca}, and in combination with DESI~\cite{DESI:2013agm}, WFIRST~\cite{spergel2015wide}, or SKA~\cite{CosmologySWG:2015ysq}. 
Fig.~\ref{fig:conts} shows the contours of $\a_s$. As the parameter $c$ approaches the gray region, the running decreases because the field value $\theta_*$ is near the hilltop of the potential.

\Eq{tensortoscalar} allows us to predict the value of tensor-to-scalar ratio $r$ where the CMB scale exits the horizon as
\begin{equation}
\label{eq:rprediction}
r \simeq \frac{8}{c \g} \rqty{1 - \frac{c^2}{4} (1 - n_{s \rm,obs})^2}.
\end{equation}
From CMB observations, there is an upper limit \cite{BICEP:2021xfz},
\begin{equation}
\label{eq:robs}
r < 0.036 \:\:\: \rm at \: 95 \% \: confidence.
\end{equation}
This provides the restriction for the pameter $c$ and the inflaton decay constant $f_\f$, 
\begin{equation}
\label{eq:rcondition}
c \gtrsim 2 \pqty{(1 - n_{s \rm,obs})^2 + 0.018 \frac{\pmass^2}{f_\f^2}}^{-\frac{1}{2}}.
\end{equation}
If $f_\f$ is much larger than $\pmass$, this restriciton is more simplified as
\begin{equation}
\label{eq:rcondtionsuper}
c \gtrsim \frac{2}{1 - n_{s,\rm obs}}.
\end{equation}
It appears to be inconsistent with the condition \eqref{eq:nscondition} so that the inflation does not work well for $f_\phi \gg \pmass$. On the other hand, in the sub-Planckian region for $f_\f$, \Eq{rcondition} can be written as
\begin{equation}
\label{eq:rcondition2}
c \gtrsim 15 \frac{f_\f}{\pmass}.
\end{equation}
This establishes a lower bound on $c$. The green region in Fig.\ref{fig:conts} represents the exclusion region imposed by upper bound on the tensor-to-scalar ratio. The green dashed line indicates the profected sensitivity of LiteBIRD\cite{Matsumura:2013aja} and CMB-S4~\cite{CMB-S4:2016ple}, corresponding to $r = 0.001$. This line is approximately given by
\begin{equation}
    c \sim 89 \frac{f_\phi}{\pmass}.
\end{equation}

The inflationary energy scale is determined by CMB normalization. The Huble parameter during the inflation is given by
\begin{equation}
    \label{eq: Hinf}
    H_\text{inf} = \pi \pmass \sqrt{\frac{\Delta^2_{\cal R, \text{obs}}}{2} r},
\end{equation}
where $\Delta^2_{\cal R, \text{obs}}$ is the observed curvature power spectrum at the CMB scale~\cite{Planck:2018jri},
\begin{equation}
\label{eq:CMBnormalizationobs}
\D^2_{{\cal R},{\rm obs}} \simeq 2.1 \times 10^{-9}.
\end{equation}
Hence, the energy scale clearly depends on the value of $r$. Moreover, it also influences the inflaton mass. If the inflaton is considered as superheavy DM in our scenario, the DM mass could offer implications for the inflationary energy scale. We discuss this further in the next subsection.

\subsection{Dark matter abundance}
In the following, we assume that $\f$ is the DM. To ensure its stability, we impose a $Z_2$ symmetry transforming $\f\to -\f$, which we refer to as dark charge conjugation symmetry, $C_{\rm dark}$. Interestingly, the potential \eqref{eq:save} manifests the symmetry.

\begin{figure}[t!]
    \begin{center}
    \includegraphics[width=81mm]{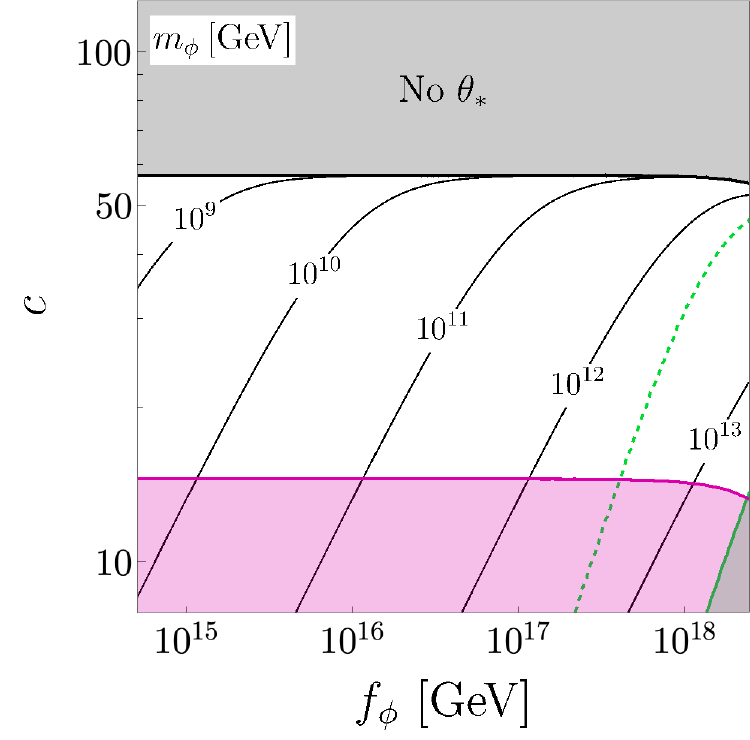}
    \includegraphics[width=81mm]{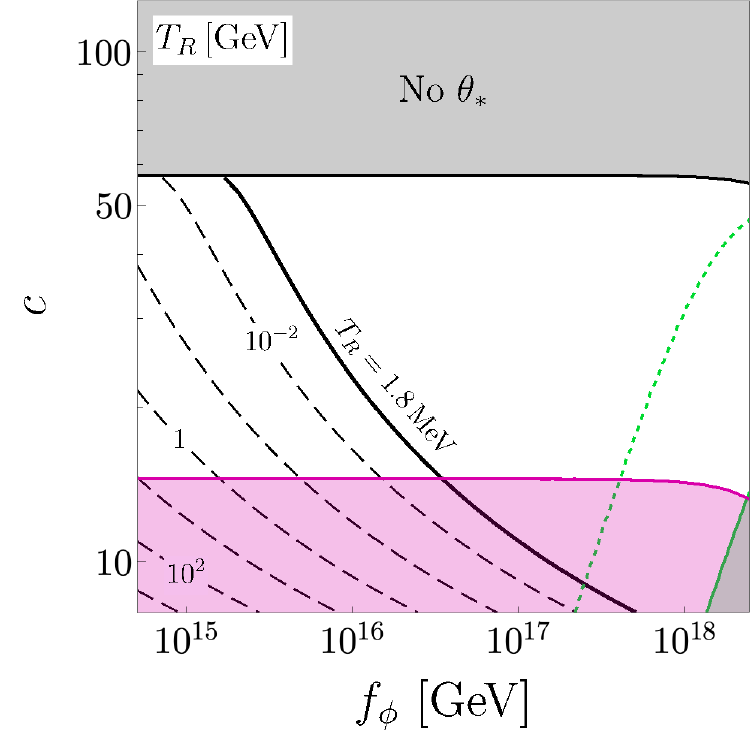}
    \end{center}
  \caption{{The left panel shows the contours of the axion mass while the right panel shows the countours of reheating temperature to explain the observed DM abundance.} Above the $T_R=1.8\MEV$ line {(bold black line)}, one needs to consider other scenarios, e.g., a second inflation driven by $\Psi$. We fix $N_{*}$ to be consistent with the thermal histry of the early universe.
  }
  \label{fig:DMabundance}
\end{figure}

Soon after the inflation, we can separate the misaligment angle into two parts as, 
\beq 
\theta = \bar{\theta}+ \d {\theta}
\eeq 
where $\bar \theta$ is the zero mode while $\d \theta$ is the non-zero mode of the field value.
The mass of $\f$ is given by 
\beq 
m^2_\phi =  \frac{\L^4}{f_\phi^2}.
\eeq 
The DM mass $m_\f$ is related to the Hubble parameter during inflation, $H_\text{inf}$, which is characterized by the shape of the potential. The Friedmann equation in the slow-roll approximation is
\begin{equation}
\label{eq:Friedmanneq}
H_{\rm inf}^2 \simeq \frac{\g \L^4}{3 \pmass^2} = \frac{c}{3} m_\f^2.
\end{equation}
Hence, Eq.~\eqref{eq: Hinf} allows us to estimate the axion mass as
\begin{equation}
\label{eq: mphiprediction}
m_\f \simeq \pmass \pi \sqrt{\frac{3 r}{2 c} \D^2_{{\cal R},{\rm obs}}} \sim 10^{10} \, \GEV \, \frac{f_\phi}{2 \times 10^{15} \, \GEV} \left(\frac{20}{c}\right)^{- \frac{3}{2}},
\end{equation}
if our scenario successfully explains the CMB observations.
Thus, measuring the axion mass can be a probe of the inflation scale. The left panel in Fig.~\ref{fig:DMabundance} represents the contours of $m_\phi$ from Eq.~\eqref{eq: mphiprediction}. For the mass range $m_\phi\gtrsim 10^{12}\,$GeV, the scenario can be probed with future measurement of the tensor-to-scalar ratio.

The DM abundance can be obtained from the misalignment mechanism~\cite{Preskill:1982cy,Abbott:1982af,Dine:1982ah}, in which the coherent oscillation of $\f$ starts when the Hubble parameter becomes comparable to $m_\f$. 
The energy of $\phi$ contributes to DM. To estimate its abundance, we need to know the value of $\theta$ at the onset of oscillation, which is determined by the inflationary dynamics. 

$\overline \theta$ can be estimated from the analysis in the previous subsection. If our scenario is to explain the CMB observations, we can also estimate the field value of $\phi$ at the end of inflation. The e-folding number from CMB horizon exit to the end of inflation is given by
\begin{equation}
\label{eq:e-folding}
N_* \simeq \frac{1}{\pmass^2}\int^{\f_*}_{\f_{\rm end}} \frac{V}{V'(\f)} d \f \simeq \int^{\h_*}_{\h_{\rm end}} \frac{c}{\sin{\h}} d \h = c \ln{\absqty{\frac{\tan{\h_*}/2}{\tan{\h_{\rm end}/2}}}},
\end{equation}
in case the inflaton slow-rolls in the negative direction. This value is related to the thermal history after inflation. In the case of single-field inflation, the e-folding number is given by
\begin{equation}
    N_* \simeq 42 + \frac{1}{6}\log \left(\frac{c}{20} \right) + \frac{1}{3}\log \left(\frac{m_\phi}{10^{10} \, \GEV} \right) + \frac{1}{3}\log \left(\frac{T_R}{10 \, \MEV}\right), \laq{efold2}
\end{equation}
where $T_R$ is the reheating temperature. If we have another field to continue the inflation, $N_*$ can be smaller. The value of $\h$ at the end of $\f$-inflation is obtained by
\begin{equation}
\label{eq:thetaend}
\h_{\rm end} \simeq 2 \arctan{\rqty{e^{-N_*/c}\sqrt{\frac{2 + c(1 - n_{s,\rm obs})}{2 - c(1 - n_{s,\rm obs})}}}}.
\end{equation}
We assume that the end of inflation is triggered by the dynamics of another field, $\Psi$. If $\f$ does not change much after $\f$-inflation, we can consider $\overline{\h} = \h_{\rm end}$.

In addition, inflationary fluctuations contribute to $\d \theta$. Since the mass is not much smaller than the Hubble parameter, we may assume that the fluctuations follow a Bunch-Davies distribution around the end of inflation. Then, the fluctuation squared has the average,
\beq f_\phi^2 \vev{\d \theta^2}= \frac{3H_{\rm inf}^4}{ 8\pi^2 m_\phi^2}.\eeq

After the inflation, this oscillation is frozen until the mass becomes comparable to the Hubble parameter due to the cosmic expansion. At the onset of the oscillation, we have the axion energy density at $H\sim m_\f$, 
\beq 
\rho_\f= m_\f n_\f= \ol{\theta}^2 f_\f^2 m^2_\f +\frac{c^2}{24\pi^2} m_\f^4.
\eeq 

We checked that the first term is usually dominant if we consider $\f$-inflation as the whole inflation period, e.g. $\bar\theta=\O(0.1)$ for an e-fold of 50. 
This would be too large to explain DM. Therefore, we consider the following two scenarios, in which we have an additional field, $\Psi$, to allow for a non-vanishing $\gamma$, in order to properly account for DM.
 One is that $\Psi$ soon starts the oscillation around its quadratic potential minimum (Late reheating scenario), the second scenario is that $\Psi$ drives another inflation (Double inflation scneario). 
Both scenarios happen depending on the dynamics of $\Psi$ followed by the two-field potential in detail.

\paragraph{Late reheating scenario:}

Let us assume that the end of inflation is triggered by $\Psi$, i.e. this is a hybrid inflation. 
Then $\Psi$ field begins to oscillate instantaneously around a quadratic potential minimum after inflation, taking over the inflation energy density, and dominating the Universe. 
The energy density of $\Psi$ is denoted by $\rho_\Psi$.
Noting that both $\rho_\f$ and $\rho_\Psi$ scale as $a^{-3}$ after the onset of axion oscillations, their ratio $\rho_\f/\rho_\Psi$ remains constant: 
\beq 
R \equiv \frac{\rho_\f}{\rho_\Psi} \simeq \frac{n_\f}{3 m_\f M_{\rm pl}^2},
\eeq 
where we used the fact that $\Psi$ dominates the Universe and {the Hubble parameter at the onset of axion oscillations is comparable to axion mass,} $m_\f=H \simeq \sqrt{\rho_\Psi/(3 M_{\rm pl}^2)}$.  The field $\Psi$ gradually decays into SM particles, forming a thermal plasma with cosmic temperature $T$ and reheating the Universe. We suppose that the reheating temperature $T_R$ is defined by the condition $\rho_r=\rho_\Psi$, where $\rho_r\equiv\frac{g_{\star}\pi^2}{30}T^4$.  
This marks the end of reheating, after which the quantity $\kappa$, defined as $\kappa \equiv \rho_\f/s$ (with $s\equiv g_{\star,s} \frac{2\pi^2}{45} T^3$), is conserved. One can relate $R$ and $\kappa$ at the end of reheating:
\beq 
\kappa= R \times \frac{3T_R}{4}.
\eeq 
In this case $N_{\rm end}$ is the end of inflation, and thus, $\bar\theta=\O(0.1)$ and dominates over $\h$. 
 The axion abundance in the present universe can be estimated as
\beq 
\Omega_\f = \kappa \frac{s_0}{\rho_c} =  \frac{T_R n_\f|_{H=m_\f} }{4m_\f M_{\rm pl}^2} \frac{s_0}{\rho_c} \simeq 0.25 \, \overline{\theta}^2 \frac{T_R}{10 \, \MEV} \left(\frac{f_\phi}{10^{15} \, \GEV}\right)^2,
\eeq 
by considering the case that the classical mode has much greater effect for the abundance than fluctuation.
{Fig.~\ref{fig:DMabundance} shows} the parameter region for the late reheating scenario, where the reheating temperature is indicated. To explain the DM abundance the thermal history is fixed. Thus e-folding number $N_*$, satisfying both \Eqs{efold2} and \eqref{eq:e-folding}, is also fixed, for a given $c$ and $f_\phi$. 
The contour set at $T_R=1.8$ MeV is a conservative bound from Big-Bang nucleosynthesis~\cite{Hasegawa:2019jsa}.\footnote{Recently, there have been many viable models for baryogenesis mechanisms with low reheating temperatures~\cite{Dimopoulos:1987rk, Babu:2006xc, Grojean:2018fus, Pierce:2019ozl, Asaka:2019ocw, Azatov:2021irb,Jaeckel:2022osh}. In addition, the decay of the heavy field $\Psi$ may serve as another source of cosmic rays, the spectrum of which could probe the reheating phase~\cite{Jaeckel:2020oet,Jaeckel:2021gah,Jaeckel:2021ert}.} 
In this scenario, the DM mass has to be smaller than $10^{12}\GEV$.

\paragraph{Double inflation scenario:}
To have even heavier inflaton we need more efficient entropy dilution. 
 In particular, the second scenario, the double inflation scenario, which has been studied to satisfy the trans-Planckian censorship conjecture~\cite{Bedroya:2019snp,Berera:2019zdd} and in the context of primordial black hole formation (e.g.~\cite{Sasaki:2018dmp}), 
is applicable to this category. 
One minimal possibility is that $\Psi$ drives a second inflation which significantly suppresses the abundance of $\f$.
We can alternatively consider thermal inflation by other fields~\cite{Lyth:1995ka, Yamamoto:1985rd}.
Although in any case the second inflationary Hubble scale should be lower than the mass of the heavy DM to suppress the fluctuations of the DM during the second inflation, 
those scenarios are highly model-dependent. Depending on the model, we can have the DM in the whole parameter region for the inflation. 
For DM with a mass larger than $10^{12} \GEV$, which is favored for the AMATERASU as well as other the highest-energy cosmic rays, we can conclude that the Universe underwent a more efficient entropy production, such as the second inflation by $\Psi$. 

\section{Phenomenology of decaying inflaton DM}
\lac{2}
The charge conjugation symmetry may be slightly broken by quantum gravity effects, as a conventional assumption for the decaying superheavy DM. Introducing explicit breaking predicts interesting phenomena for cosmic rays. 
{In our scenario, the DM is a singlet scalar particle. Thus, it is important to study its decay while taking into account the chirality suppression effect.}

\subsection{DM decay formula}

Let us discuss the general aspects of DM decay.
\lac{pre}
To calculate DM decay, we can consider the following two components:
\beq
\F_X= \F_X^{\text{extra}}+\F^{\rm MW}_{X}.
\eeq
Here, $X$ denotes the particle produced in DM decay. The first term represents the extragalactic component, which is isotropic. 
We neglect this component for simplicity (see, e.g., Ref.~\cite{Murase:2015gea} for the extragalactic contribution). 
For our purpose to explain the KM3-230213A event it is not important because it only contributes to a subdominant lower-energy component rather than the peak (see Fig.~\ref{fig:Hll}).

The second term, which is our focus, represents the component originating from the Milky Way. For our purpose, we consider the all-sky average intensity~\cite{Murase:2012xs,Das:2023wtk}, 
\begin{eqnarray}
\Phi_{X}^{\rm MW}[E_X]&=&\frac{1}{4\pi}
\int ds \int_{\D\Omega=4\pi}  d \Omega \frac{e^{-o[s, \Omega ] s}}{4\pi s^2}   \left(\frac{\t^{-1}  \rho^{\rm MW}_\phi(s,  \Omega )}{m_\phi}\right) \, s^2\, \frac{d N_{\f}}{dE_X}\nonumber\\
&=& \frac{e^{-o[s, \Omega ] s}}{4\pi m_\phi \t} \bar{D}^{\rm MW}_{\D \Omega=4\pi} \, \frac{d N_{\f}}{dE_X},
\end{eqnarray}
where $s$ is the line-of-sight distance, 
$\rho^{\rm MW}_\f$ and $\frac{d N_{\f} }{dE_X}$ represent the DM density distribution and the $X$ spectrum from a DM decay at rest, respectively. 
$\D\Omega$ is the solid angle of interest.
We neglect the DM motion because the detector's energy resolution is insufficient to resolve the Doppler shift of the DM around us. 
$o$ is the (averaged) optical depth for absorption, e.g.,~Ref.~\cite{Cirelli:2010xx} . We take 
$
o\simeq 0
$
unless otherwise stated. $\tau$ denotes the DM lifetime.

We introduced the so-called $D$-factor and its average value is
\beq
\bar{D}_{\D \Omega=4\pi}=\frac{1}{4\pi} \int_{\D \Omega} d \Omega \int ds \rho^{\rm MW}_{\f} (s, \Omega) .
\eeq

To calculate the $D$-factor, we should introduce a specific form of the DM density distribution. For our galaxy, the NFW profile~\cite{Navarro:1996gj} is usually adopted,
\beq\laq{NFW}
\rho_{\rm NFW}(r)=\frac{{\rho_0}}{\frac{r}{r_s}\left(\frac{r}{r_s}+1\right)^2}.
\eeq
This can be fitted from Gaia DR2~\cite{Cautun:2019eaf}, with the parameters $\rho_0\approx 0.46 \, \GEV/\rm cm^3$, $r_s\approx 14.4 \, \text{kpc}$.\footnote{Alternatively, we also adopt the Einasto profile~\cite{Einasto:1965czb,Navarro:2003ew}.
$
\rho_{\rm Einastro}(r)= \laq{Einastro}\rho_0 e^{-(\frac{r}{h})^{1/n}},
$
{that is fitted} from Gaia DR3~\cite{Jiao:2023aci}{, with the parameters} $\rho_0\approx 0.76 \GEV {\rm cm}^{-3}, n\approx 0.43, h\approx 11.41$ kpc. This does not change the numerical results much. }
{We introduce the distance from the center of our galaxy, $r$, which satisfies the relation:} 
$$
r^2=s^2+ r^2_{\odot}-2 r_{\odot} s \cos\theta,
$$ 
with $r_{\odot}\approx 8.2\rm\, kpc$ is the distance between the solar system and the center of the galaxy.

\subsection{DM decay spectra}

To obtain {the flux of particles $X$ in DM decay, $d N_{\f}/d E_X$}, we require a particle theory. Since the DM is a spin-zero scalar field and is much heavier than the weak scale, the main decay channels, via operators with a dimension no larger than 5, are expected to be,
\beq
\phi \to H \bar q q,\; \bar H \bar l l,\; \bar H H,\; gg,\; AA,\; BB,
\eeq
with $g$, $A$, and $B$ being the gauge fields of $\SU(3)_c$, $\SU(2)_L$, and $\U(1)_Y$, respectively (depending on the coupling). Now, we use
the notation in the symmetric phase. 
$q, l, \AND H$ denotes a quark, a lepton and a (charged) Higgs boson, respectively, with bar denoting the anti particle. 
Since we consider $\phi$ to be a pseudoscalar, the decay into two Higgs bosons must occur via CP violation and can be highly suppressed depending on the model-building. We also note that the decay of $\phi$ into two fermions is forbidden by chirality arguments. In the context of indirect detection, 
we typically assume that the decay is a two-body process. The three-body decays and $\f \to gg$ have not been well studied. Hence, in this paper, we study the possibilities $\f \to H\bar qq$, $H \bar l l$, and $gg$. In any case, cascade decay and fragmentation determine the branching fraction of the final asymptotic-state particle. For the $\f \to gg$ channel, we adopt {\tt HDMSpectra}\footnote{We found a strange peak, which would presumably be unphysical, near $x=1$ in the fragmentation function for photon production from $W$ and $Z$ bosons. This peak originates from the transverse modes. However, in our scenario, the DM $\phi$ decays exclusively into the longitudinal mode due to the equivalence theorem. As a result, this strange peak is absent in our case.} \cite{Bauer:2020jay}, which can estimate the spectrum for DM two-body decays. In the three-body decay case, we first estimate the primary decay process to determine the fraction of the SM particles, and then use {\tt HDMSpectra} to estimate the subsequent cascades. 
The details of the former are described in Appendix \ref{app:1}. More precisely, in {\tt HDMSpectra}, the two-body DM decay $\f \to Y\bar Y$ provides the cascading of a monochromatic spectrum with energy $E_Y=m_\phi/2$ into the final particle $X$ with energy $E_X$, i.e., $D^X_Y(E_X/E_Y)$. Here, the fragmentation function $D^X_Y(x)$ represents the probability that an initial particle $Y$ produces the final particle $X$ carrying a momentum fraction $x$.
Then, from the fractional distribution $f_Y(E_Y)$, estimated in Appendix \ref{app:1}, we obtain
\begin{equation}
    \frac{{d} N_\f}{{d} E_X}  =\sum_Y \int^{\infty}_{E_{X}} \frac{{d} E_Y}{E_Y} f_Y(E_Y) D^X_Y(E_X/E_Y).
\end{equation}
wherer $d N_\phi/d E_X$ is the spectrum of $X$ from the rest $\phi$ decay.

\subsection{Results}

We consider three final states, $\phi\to  H \bar{q} q$, $\phi\to  H \bar{l} l$, and $\phi\to  gg$.

\paragraph{$\phi\to  H \bar{q} q$:}
Let us first discuss the case 
\beq 
\label{eq:3body}
\phi\to  H \bar{q} q,
\eeq 
with $H$ being the SM Higgs field. We note that $\phi$ decays into fermions predominantly via three-body processes due to chiral suppression. 

\begin{figure}[!t]
\vspace{-3mm}
    \begin{center}
        \hspace{-10mm}
        \includegraphics[width = 80mm]{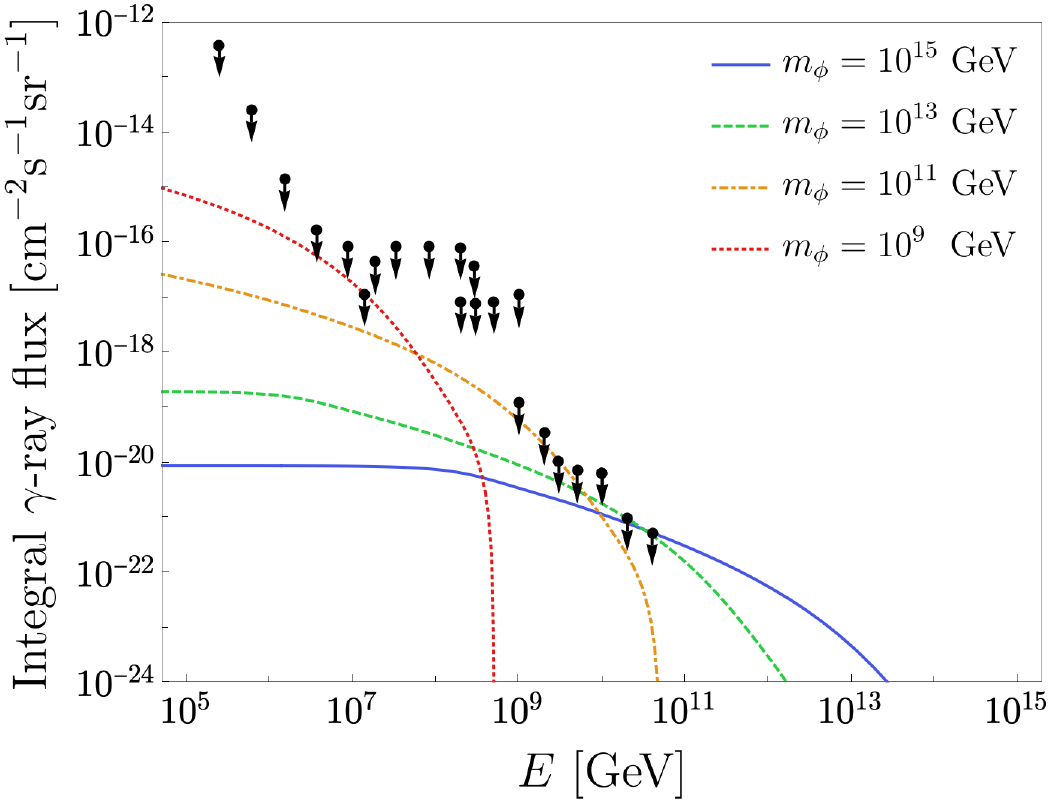}
        \hspace{10mm}
        \includegraphics[width = 63mm]{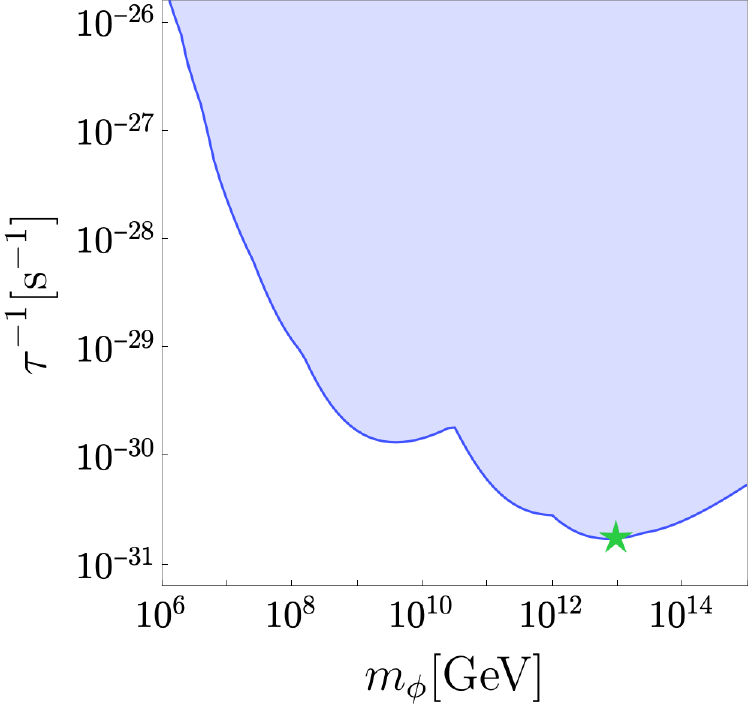}
        \includegraphics[width = 80mm]{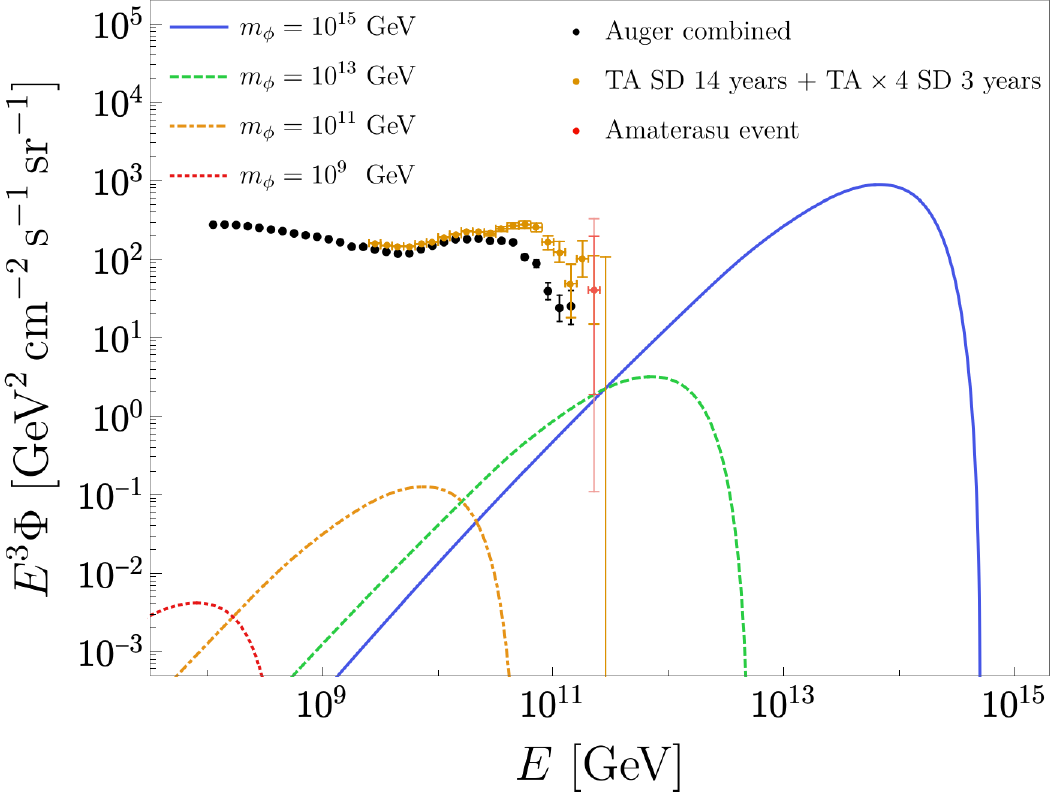}
          \includegraphics[width = 80mm]{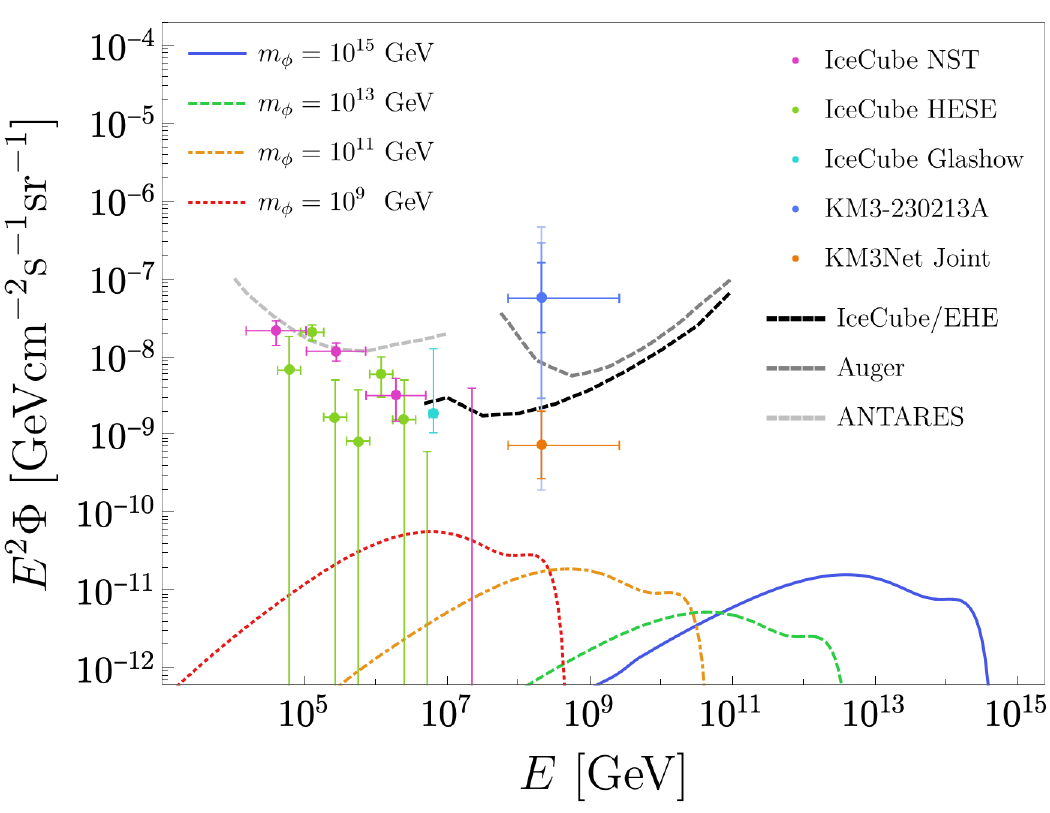}
    \end{center}
          \vspace{-7mm}
  \caption{[\textit{Left top}] The {\it integrated} photon fluxes for the decay channel, $\phi \rightarrow H q \bar{q}$, with $m_\phi = 10^9, \, 10^{11}, \, 10^{13}, \,10^{15} \, \GEV$, that just sastify the photon limits. The limits from KASCADE, KASCADE-Grande~\cite{KASCADEGrande:2017vwf}, and Auger~\cite{Savina:2021cva, PierreAuger:2022uwd, PierreAuger:2022aty} are also shown. [\textit{Right top}] The upper limits of the DM decay rates. 
  The green star indicates the grean dashed lines ($m_\phi = 10^{13} \GEV$) in other figures relevant to the AMATERASU event.
[\textit{Left bottom}] The corresponding averaged fluxes of the CR {$p+\bar p+n+\bar n$}. The black data points from Auger are taken from~\cite{PierreAuger:2019phh}. The colored ones are from the TA experiment~\cite{Sagawa:2022glk, TelescopeArray:2018xyi}. The red point corresponds to the AMATERASU particle~\cite{TelescopeArray:2023sbd}. 
 Following Poisson statistics, we express the uncertainty up to $3\sigma$.
  [\textit{Right bottom}] The fluxes of $\frac{1}{3}\sum \nu+\bar\nu$. {The blue and orange points represent the best-fit data point by KM3NeT and the IceCube-KM3Net joint data point, respectively~\cite{KM3NeT:2025npi}. The other data are shown as NST (purple points)~\cite{Abbasi:2021qfz}, HESE (green points)~\cite{IceCube:2020wum}, and Glashow resonance event(light blue points)~\cite{IceCube:2021rpz}. Also, the dashed lines represent upper limits from IceCube-EHE (90\% CL~\cite{IceCube:2025ezc}), Auger (90\% CL~\cite{PierreAuger:2023pjg}), and ANTARES (90\% CL~\cite{ANTARES:2024ihw}).} 
  The error bars for the KM3NeT data indicate the $1,2,3 \sigma$ uncertainties.
  }
  \label{fig:Hqq}  \label{fig:Hqqnucleon}
\end{figure}

For concreteness, let us consider an interaction of the type, 
\beq 
{\cal L}\supset - \frac{\phi H\bar{u} \hat{P}_L Q}{{M_Q}}+h.c.,
\eeq 
where {$M_Q$} is a higher-dimensional scale, $Q$ is the quark doublet, and $u$ is the right-handed quark. According to the gauge invariance,  
there are six types of decay channels in the broken phase: $\f\to h \bar u_L u_R$, $\f\to W_- \bar d_L u_R, \f \to Z \bar u_L u_R$, $\f\to h u_L \bar  u_R$, $\f\to W_+  d_L \bar  u_R, \f \to Z  u_L \bar u_R$. The contributions with certain wight according to the equivalence theorem can be found in Appendix.\,\ref{app:1}.

As discussed in Ref.~\cite{Das:2023wtk}, the most stringent constraint for superheavy DM decay comes from $\gamma$-ray observations. Thus, we first study 
{the flux of photons} to determine the bound on {the DM lifetime} $\tau$. 
We adopt a conservative approach to estimate the constraint on photons.

This approach allows us to compare with the $2\sigma$ data from KASCADE, KASCADE-Grande, and Auger~\cite{KASCADEGrande:2017vwf,Savina:2021cva, PierreAuger:2022uwd, PierreAuger:2022aty} as shown in the left top panel of Fig.~\ref{fig:Hqq}. This {observational} bound is translated into a lower limit on the DM lifetime. 
{The right top panel of Fig.~\ref{fig:Hqq} shows the constraints on the DM decay rate  (the inverse of the lifetime) as a function of the DM mass.}

We {also} estimate the averaged flux of nucleons, including protons, antiprotons, neutrons, and antineutrons. If the CRs are composed of charged particles including protons and antiprotons, the magnetic fields largely affect the trajectory. We emphasize that if the superheavy DM can decay into protons, it can also decay into neutrons. The highest-energy neutron from the DM decay in nearby galaxies or our galaxy may not decay into a proton before reaching Earth. This is because the lifetime (or decay length) with the Lorentz factor is 
\beq
\tau_n\approx \frac{E_n}{m_n} 880 \ {\rm s}\simeq 0.91 \,{\rm Mpc}~\frac{E_n}{10^{11}~\GEV}.
\eeq
In general, it is difficult for air-shower detectors to distinguish between (anti)protons and (anti)neutrons. However, with this neutron contribution, anisotropy toward the Milky Way and other nearby galaxies would be enhanced, which may be tested by future bigger UHECR data with more statistics.  

By applying the upper limit on $\tau^{-1}$ derived from observational constraints on the photon flux,
the results for {$p+\bar p+n+\bar n$} are shown in the left bottom panel of 
Fig.~\ref{fig:Hqqnucleon} 
The data points are taken from the UHECR spectrum data~\cite{PierreAuger:2019phh} and the TA data~\cite{Sagawa:2022glk, TelescopeArray:2018xyi}.  For illustrative purposes, we show the results $m_\f = 10^{9,11,13,15} \, \text{GeV}$.

In the bottom right panel, we denote the single-flavor neutrino fluxes (averaged neutrino antineutrino flux by the 3 generation) from the same decay channel. The fluxes are from the DM decay with the same decay rate as the ones shown in the left figures.  
Other datasets are depicted as NST (purple points)~\cite{Abbasi:2021qfz}, HESE (green points)~\cite{IceCube:2020wum}, and the Glashow resonance event (light blue points)~\cite{IceCube:2021rpz}. The dashed lines indicate 90\% CL upper limits from IceCube‑EHE~\cite{IceCube:2025ezc}, Auger~\cite{PierreAuger:2023pjg}, and ANTARES~\cite{ANTARES:2024ihw}.

\paragraph{$\phi\to  \bar H \bar{l} l$:}

\begin{figure}[!t]
    \begin{center}
        \hspace{-10mm}
        \includegraphics[width = 80mm]{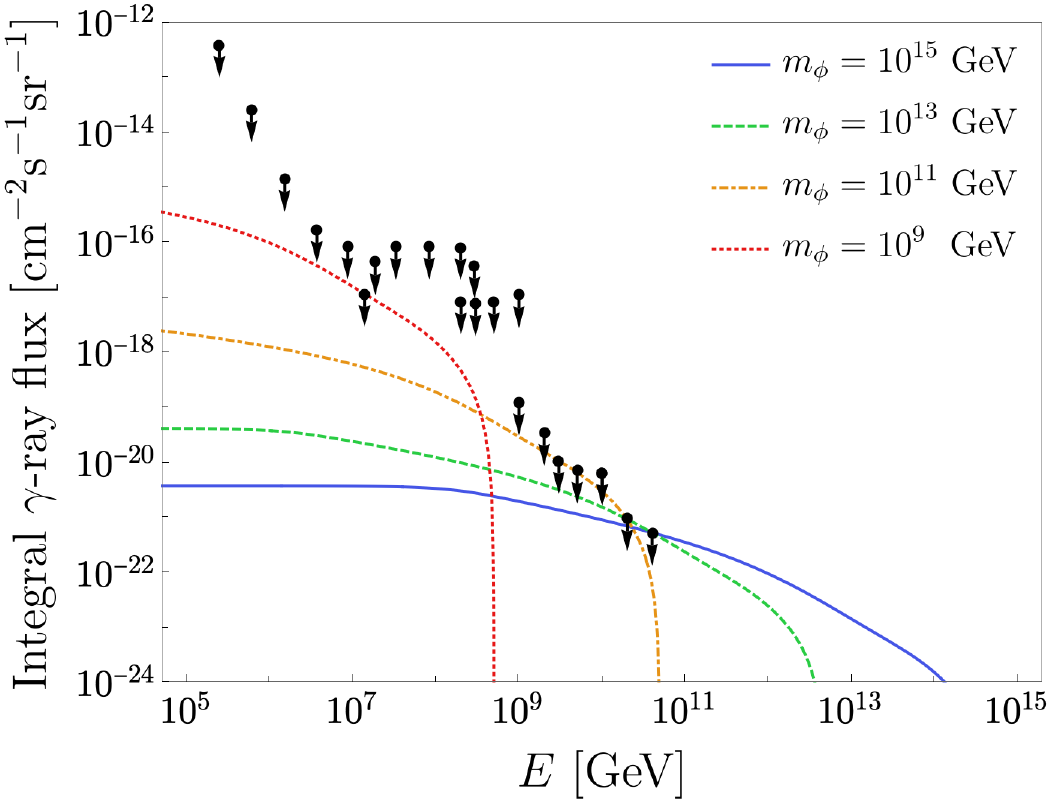}
        \hspace{10mm}
        \includegraphics[width = 63mm]{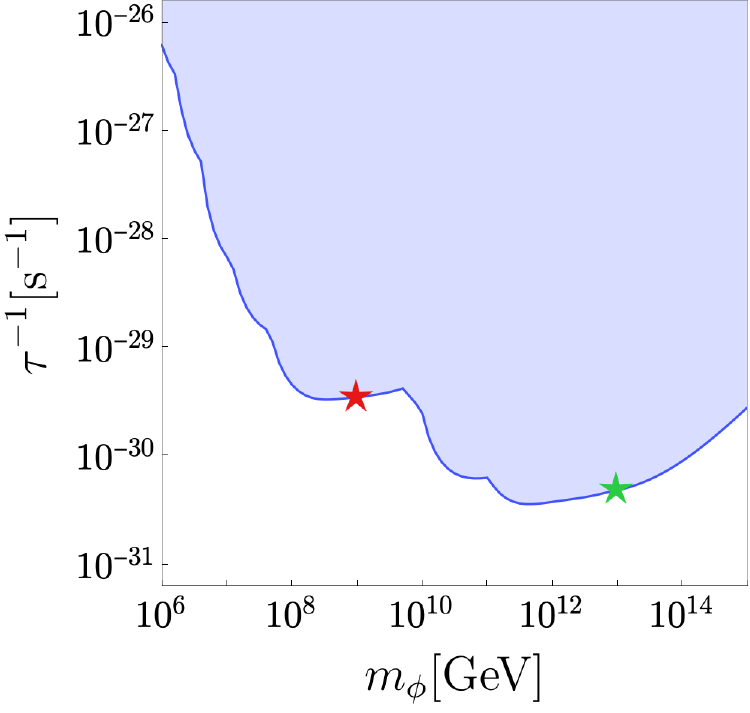}
        \includegraphics[width = 80mm]{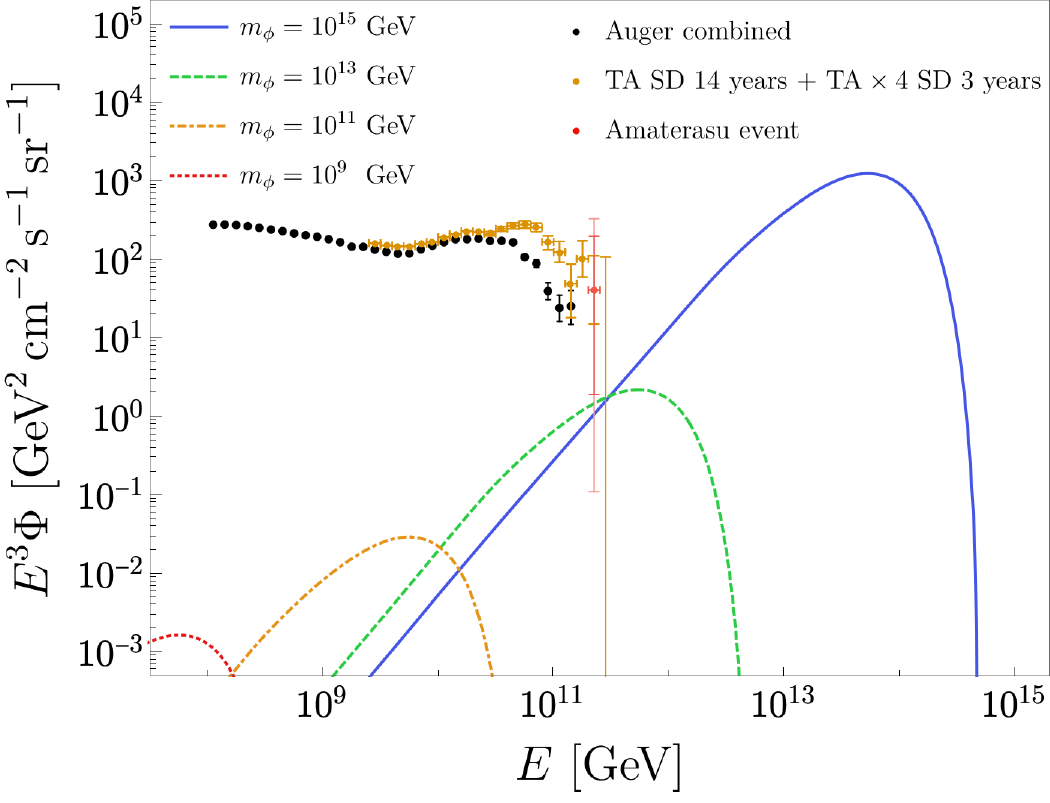}
        \includegraphics[width = 80mm]{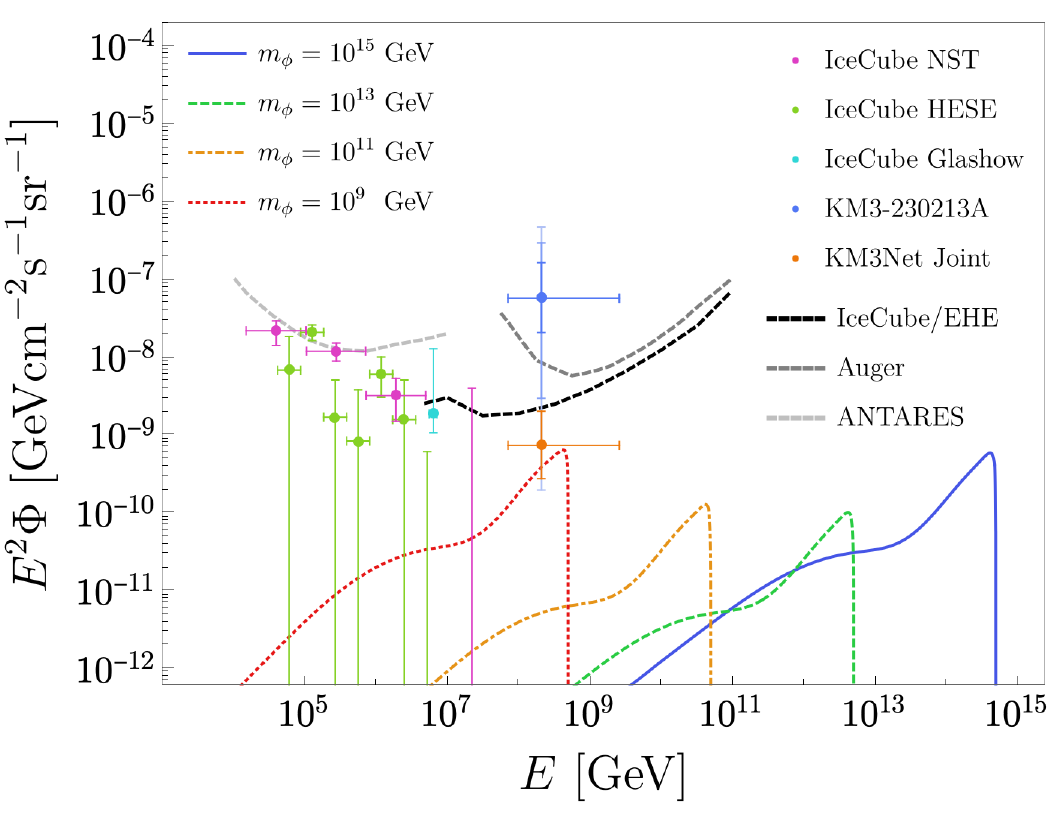}
    \end{center}
    
    \caption{
    Same as Fig.~\ref{fig:Hqq}  but assuming the channel,  $\phi \rightarrow \bar{H} \bar{l} l$. In the right-top panel, the red star corresponds
    to the KM3-230213A event in addition to those for the AMATERASU event (green star).
    }
    \label{fig:Hll}
\end{figure}

Next, let us consider an interaction of another type:
{\begin{equation}
    \mathcal{L} \supset - \frac{\phi \bar{H} \bar{e}_R \hat{P}_L L}{{M_L}}+h.c.,
\end{equation}}
Similar to the case of $\phi \to H \bar{q} q$, there are six types of decay channels in the broken phase:  $\f \to h \bar e_L e_R$, $\f \to W_+ \bar \nu_L e_R$, $\f \to Z \bar e_L e_R$, $\f \to h e_L \bar  e_R$, $\f \to W_-  \nu_L \bar  e_R$, and $\f \to Z e_L \bar e_R$. We can calculate these as well, following the procedure in Appendix \ref{app:1} (with the color factor removed). A similar 
results are shown in Fig.~\ref{fig:Hll}. One can see that the nucleon flux in this case is further suppressed, especially at low energies.

\paragraph{$\phi\to g g$:}
Let us consider
\begin{figure}[!t]
    \begin{center}
        \hspace{-10mm}
        \includegraphics[width = 80mm]{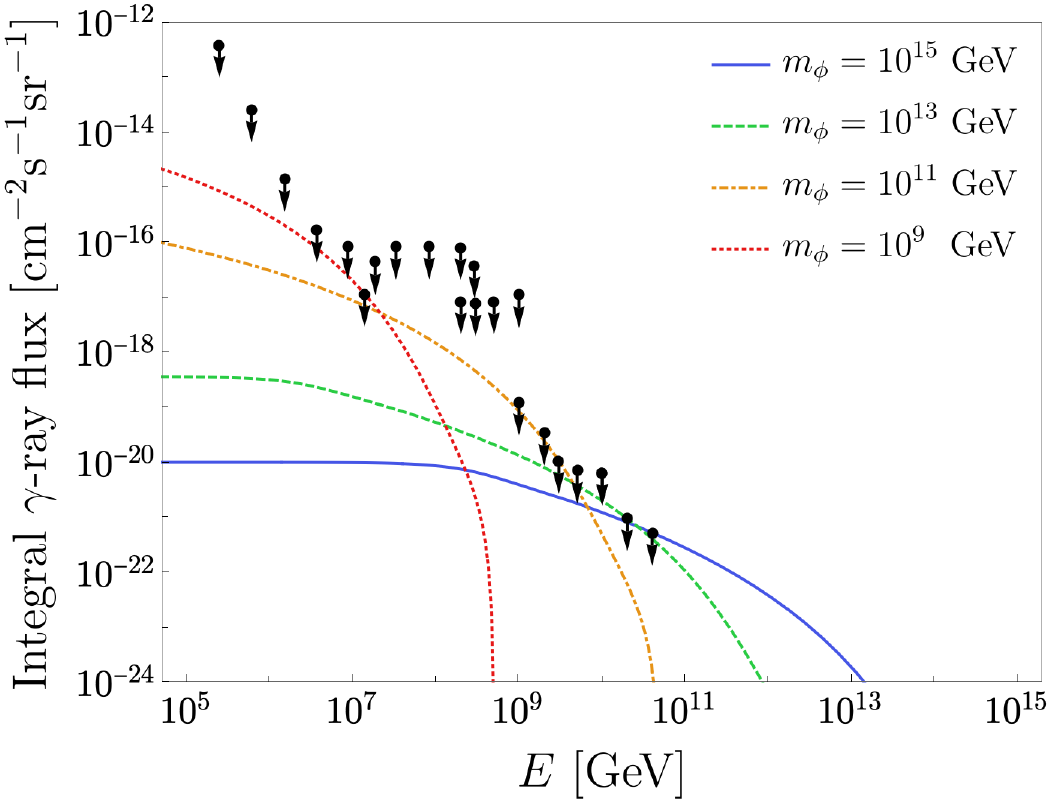}
        \hspace{10mm}
        \includegraphics[width = 63mm]{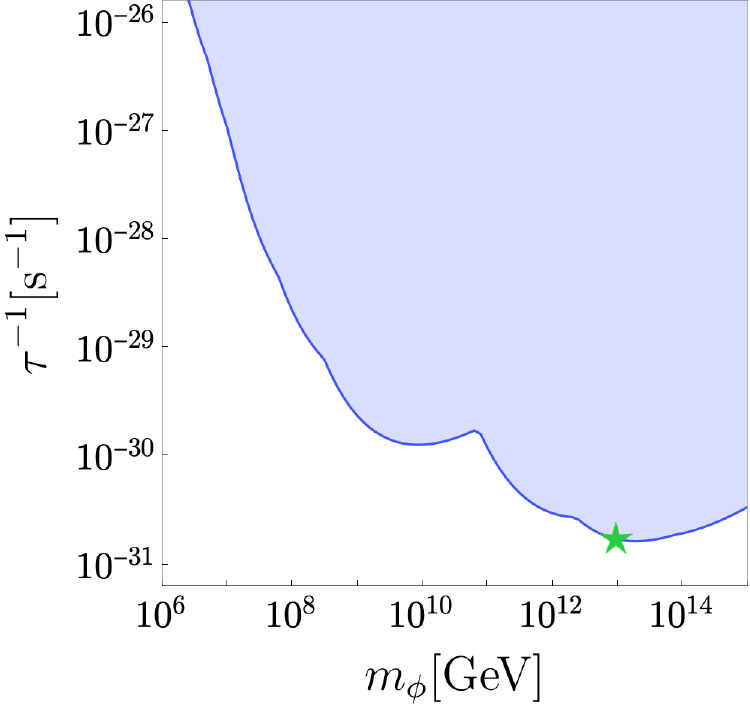}
        \includegraphics[width = 80mm]{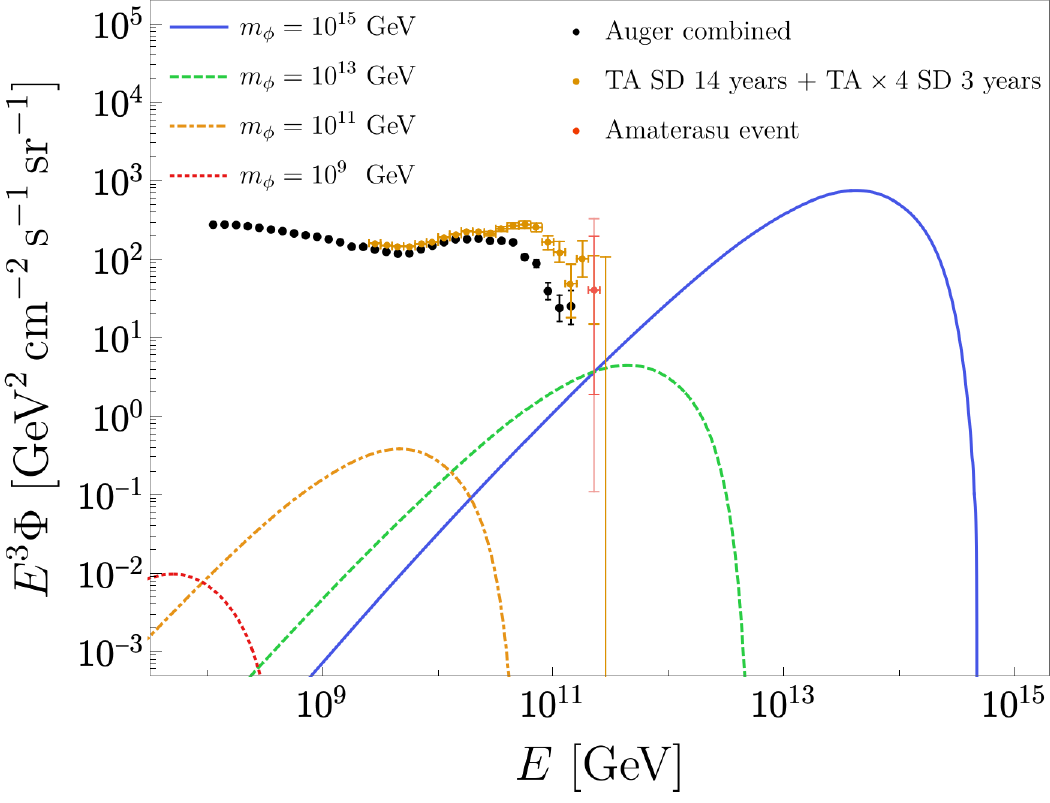}
        \includegraphics[width = 80mm]{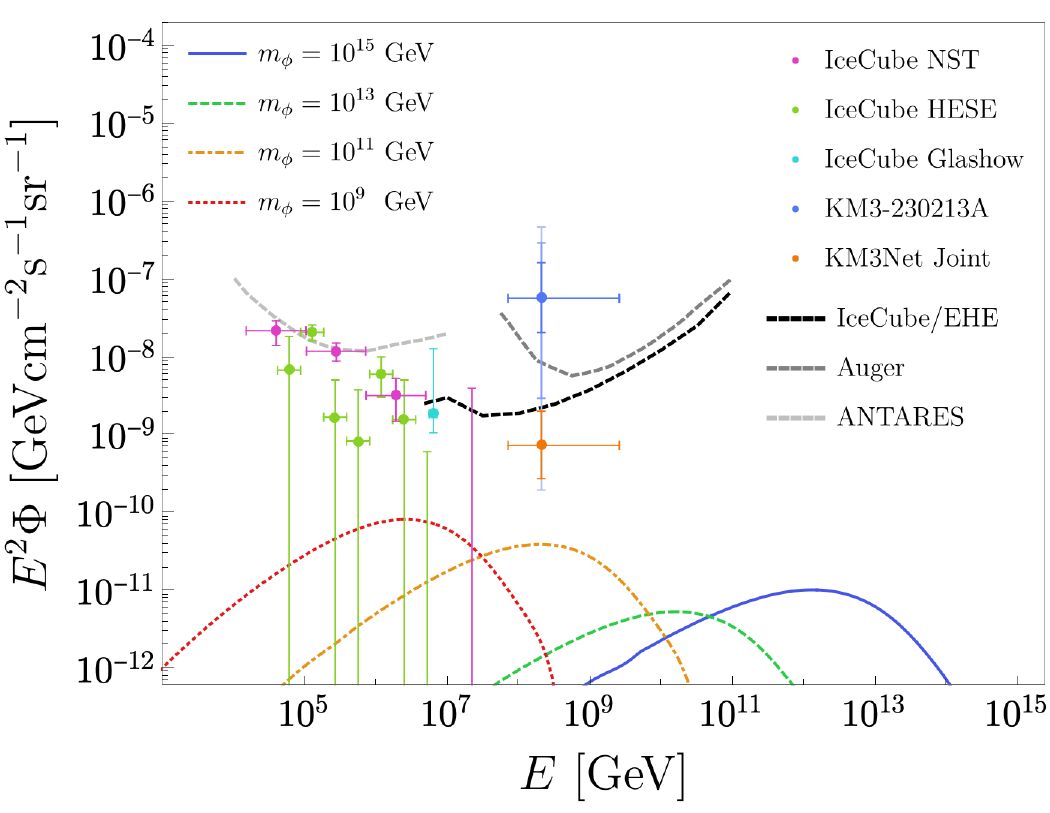}
    \end{center}

    \caption{
    Same as Fig.~\ref{fig:Hqq} but assuming the channel,  $\phi \rightarrow gg$.
    }    
    \label{fig:gg}
\end{figure}

\beq
{\cal L}\supset \frac{\phi}{{M_G}}G_{\mu\n}G^{\m\n},  \frac{\phi}{{\tilde{M}_G}}G_{\mu\n}\tl G^{\m\n}
\eeq
where $G$ is the gluon field strength and $\tl G$ denotes the dual of $G$. 
The results are shown in Fig.\ref{fig:gg}. Notably, the bottom left panel of Fig.\ref{fig:gg} indicates a maximal nucleon flux comparable to that in Fig.~\ref{fig:Hqq}.

The recently observed AMATERASU event~\cite{Fujii:2024sys} can be explained within the 2$\sigma$ level for the $\phi\to gg$ or $\phi\to Hqq$. For $\phi\to \bar H\bar l l$ it is within the 3$\s$ level.  The AMATERASU event originates from a local void where there are no known sources for the acceleration of the highest-energy cosmic rays. This is consistent with our scenario, especially since the event direction is near the galactic center (and also close to the Hercules dwarf galaxy). If this is the case, 
we predict that the DM mass is heavier than $10^{13}\GEV$ (see Figs.~\ref{fig:Hqqnucleon}, \ref{fig:Hll}, and \ref{fig:gg}) and 
the tensor-to-scalar ratio and running of the spectral index will 
prove our model in the future measurement (see Fig.~\ref{fig:conts}). 
{However,} a second inflation may be necessary to explain the DM abundance (see Fig.~\ref{fig:DMabundance}).

On the other hand, for the KM3-230213A event, the DM mass is preferred to be around $10^{9}\GEV$ when $\f$ decays into letpons (see the right bottom panel of Fig.\ref{fig:Hll}). This can be consistent with the parameter region of the late reheating scenario. 
We note that the KM3-230213A  event is in tension with the other observations such as IceCube and Auger. However, combining them gives a lower flux~\cite{KM3NeT:2025ccp}, which could be explained in our scenario within the $1\sigma$ level via the leptonic decays.

\section{Particle model for the highest-energy cosmic rays with suppressed $\gamma$-ray signals}
\lac{PM}
So far, we have discussed the possibility that the heavy DM decays into SM particles and attempted to explain some of the highest-energy cosmic rays, including the AMATERASU event but find that we can only explain it within $2\sigma$ level due to the severe multi-messenger $\gamma$-ray bound. 
Before concluding this paper, we propose a particle model that allows 
{to give better fit of the AMATERASU event and explain a significant part of the} highest-energy nucleons. Indeed, given that the inflaton mass is predicted to be high in our scenario, the Beyond SM, if exists, should matter if the mass scale is lower than inflaton.

Let us consider the scenario where heavy DM decays into dark sector particles. Then, a dark particle among the decay products further decays into a nucleon and additional dark particles. 
For concreteness, consider the following decay chain:
\beq
\f \to N_{\rm dark} + X_{\rm dark}, ~~~ N_{\rm dark} \to n + Y_{\rm dark}.
\eeq
Here and hereafter, the subscript ``dark" denotes the dark components which do not carry SM gauge charges. $X_{\rm dark}$ and $Y_{\rm dark}$ are dark particles or a collection of dark particles and $n$ represents the neutron. To suppress the production of other SM particles, we make three assumptions:
\begin{itemize}
    \item{\bf (1) Small $Q$-value for the second reaction:} The last reaction has a $Q$-value not much larger than $\GEV$, so that the decay can be accurately estimated using a hadronic description rather than a quark-level description.
    \item{\bf (2) Baryon number symmetry:} We assume baryon number conservation to forbid proton decay; thus, $N_{\rm dark}$ and/or $Y_{\rm dark}$ carries baryon number.
    \item{\bf (3) Dark symmetry:} We assume a dark symmetry under which $N_{\text{dark}}$ is charged, and consequently, $X_{\text{dark}}$ and $Y_{\text{dark}}$ have the opposite and same charge since the real scalar $\f$ is not charged.
\end{itemize}

Assumptions (2) and (3) suppress reactions involving other unwanted SM particles, such as $N_{\rm dark}\to n + \gamma$, $N_{\rm dark}\to p + Y_{\rm dark}$, or $N_{\rm dark}\to \nu/e + Y_{\rm dark}$. 
Some processes like $N_\text{dark} \to n + \gamma + Y_\text{dark}$  may still occur without violating the imposed symmetries. However, they are suppressed as higher-order effects in renormalizable models, as we will see.
Particle models that give rise to these reactions can be constructed, as detailed in the following.

\subsection{A simple model}
Before constructing a more natural model, let us study a minimal model that realizes this idea. The Lagrangian in the hadronic picture is given by 
\beq
{\cal L}\supset \e_{\text{DM}} \f i\bar N_{\rm dark} \gamma_5 N_{\rm dark} - g_{\rm dark} \bar N_{\rm dark} n \pi_{\rm dark}.
\eeq
Here, $\e_{\text{DM}}$ represents a small violation of the symmetry that stabilizes the DM {and $g_\text{dark}$ is a coupling constant for neutron-dark sector interaction. The first term} can naturally be small enough to yield a sufficiently long lifetime for the DM. Therefore, we focus on the {DM} lifetime $\tau$ rather than $\e_{\text{DM}}$. We identify $X_{\rm dark}=\bar{N}_\text{dark}$ and $Y_{\rm dark}=\pi_{\rm dark}$, where $\pi_{\rm dark}$ is a dark real scalar (not a pseudoscalar in this context). In addition, we assume that $\bar{N}_{\rm dark}$ and $\pi_{\rm dark}$ carry a certain $Z_2$ symmetry, which forbids terms such as $\bar N_{\rm dark}\gamma_5 n \pi$, or $\bar N_{\rm dark}F_{\m\n}[\gamma^\m, \gamma^\n] n$, etc., where $\pi$ would be the pion and $F_{\m\n}$ is the photon field strength {in the SM}. 
The two-body decay width {of $N_\text{dark}$} is given by 
\begin{equation}
  \Gamma = \frac{g_{\rm dark}^2}{16 \pi m_{N,\rm dark}^3} \lambda (m_{N,\rm dark}^2, m_{n}^2, m_{\pi,\rm dark}^2) \big[ (m_{N,\rm dark} + m_{n})^2 - m_{\pi,\rm dark}^2 \big].
\end{equation}
where 
\begin{equation}
    \lambda(x,y,z) = \sqrt{x^2 + y^2 + z^2 - 2xy - 2xz - 2yz},
\end{equation}
is the square root of the K\"{a}ll\'{e}n function. By expanding in the $Q$ parameter, defined as
\beq
Q {\equiv} m_{N,\rm dark} + m_{n} - m_{\pi,\rm dark},
\eeq
we obtain
\beq
  \Gamma \simeq \frac{g_{\rm dark}^2}{2\sqrt{2}\pi} \frac{m_n\sqrt{m_n Q m_{\pi,{\rm dark}} m_{N,\rm dark}}}{m_{N,\rm dark} }.
\eeq
Since $Q \lesssim \GEV$ for our hadronic description to hold, we find that this decay width is maximized when $m_{N,{\rm dark}} = \mathcal{O}(\GEV).$ Then, we obtain $\Gamma \sim 0.1\, g_{\rm dark}^2\, m_{N}.$ Thus, for a boosted $N_{\rm dark}$ with energy $E_{N,\rm dark}$, the decay length is given by\footnote{We do not consider the scenario where $l_{\rm decay}\gg 100{\rm Mpc}$ (the scale relevant for the GZK cutoff) in this paper, or where the decay rate of $\f$ is higher than previously discussed. In such a scenario, $\f$ decay produces dark radiation, and the large amount of dark radiation decays into neutrons with a suppressed probability to explain the flux of the highest-energy cosmic rays. 
{Then}, the extragalactic component of $N$ would dominate the production of the highest-energy cosmic rays (see, e.g., Ref.~\cite{Jaeckel:2021ert}). However, the photomeson production must produce too many $\gamma$ and $\nu$, which are {strongly} constrained {by observations.}
The higher-order decay processes of $\f$ in the Milky Way galaxy would involve $\gamma$ and $\nu$ in the UV model. For larger decay length, those fractions compared to the neutron from the decaying $N_{\text{dark}}$ in the small range, are also enhanced and are constrained.  For the sterile neutrino case~\cite{Narita:2025udw}, this is higher order effect is not important because of the very weak renormalizable interaction of the sterile neutrino. }
\beq
l_{\rm decay}\sim 0.6{\rm \,kpc}\frac{E_{p,\rm dark}}{10^{12}\GEV} \left(\frac{10^{-11}}{ g_{\rm dark}}\right)^2.
\laq{ldecay}
\eeq
{Therefore}, for $ g_{\rm dark}\gtrsim 10^{-12}$, the $N_{\rm dark}$ is almost completely converted into a neutron by the time it reaches Earth.

\paragraph{UHECR spectra}

To illustrate the spectrum, we note the energy of the resulting neutron
\beq
E_n\simeq \frac{E_{N,\rm dark}}{2m_{N,\rm dark}}\Bigl(\sqrt{p_{\rm rest}^2+m_n^2} + p_{\rm rest} \cos \theta_{\rm rest}\Bigr), 
\eeq
where 
$$
p_{\rm rest}=\frac{\sqrt{m_n^4-2 m_{N,\rm dark}^2 \left(m_{\pi,\rm dark}^2+m_{n}^2\right)+\left(m_{\pi,\rm dark}^2-m_{n}^2\right)^2}}{2 m_{N,\rm dark}}
$$
is the momentum in the $N_{\rm dark}$ rest frame, and $\theta_{\rm rest}$ is the polar angle with respect to the boost direction in the rest frame. Due to rotational invariance, $\cos(\theta_{\rm rest})$ is uniformly distributed in $[-1,1]$. Thus, the spectrum is given by
\beq
\F=\text{const.}
\quad 
{\rm for } \quad E_n(\cos \theta_{\rm rest} =-1)<E<E_n(\cos \theta_{\rm rest} =1),
\eeq
and  zero otherwise{, which results in}
a sharp UV and IR cutoff. For a small $Q$ expansion, we 
{obtain the energy range where the highest-energy cosmic ray appears as follows:}
$$ 
\frac{E_{N,\rm dark}}{2m_{N,\rm dark}} \Bigl(m_n- \sqrt{2Q} \sqrt{\frac{m_n m_{\pi,\rm dark}}{m_{N,\rm dark}}}\Bigr) < E < \frac{E_{N,\rm dark}}{2m_{N,\rm dark}} \Bigl(m_n+ \sqrt{2Q} \sqrt{\frac{m_n m_{\pi,\rm dark}}{m_{N,\rm dark}}}\Bigr).
$$
{Thus,} this {mechanism} can provide a good fit of the AMATERASU event. Also it may provide a  good fit for various data for the highest-energy cosmic-ray spectrum while without affecting the spectrum below $10^{10}\GEV$, where HECRs may not be the  (anti)nucleon, or above $10^{12}\GEV$, where no CR is observed.

\paragraph{UV Completion} Given the large DM mass, to justify the effective theory at the $\GEV$ scale, we consider the UV completion of the model.  
Since the interaction can be obtained from a higher-dimensional quark Lagrangian, {we can estimate the following interaction:}
\beq
\frac{\bar N_{\rm dark} u d d \pi_{\rm dark}}{M_q^3} \to \frac{\L^{3}_{\rm QCD}}{M_{q}^3} {\bar{N}}_{\rm dark} n \pi_{\rm dark},
\eeq
with $u \AND d$ again representing the right-handed up and down quarks, respectively. It implies
that $g_{\rm dark}\simeq  {\L^{3}_{\rm QCD}/M_{q}^3}$. This requires $M_q < \O(\TEV)$ to achieve a sufficiently high decay rate, which in turn suggests the existence of light charged particles for the UV completion. One example is\footnote{In this UV completion, we also have  $\f \to N_{\rm dark} u \f_u^*$, etc. The final state cascades induce neutrinos and $\gamma$ rays, but these are suppressed by phase space and additional couplings, rendering the flux more than two orders of magnitude smaller than that of $n$.} 
\beq 
{\cal L}_{\rm UV}\supset -\tl y_N \bar{N}_{\rm dark} u \phi^*_{u} - A_{\phi}\f_u \tilde{\phi}^*_u \pi_{\rm dark} - \tl{y}_{\f} \tilde{\phi}_u d d,
\eeq
$\phi_u, \tl\phi_u$ are di-quarks which are scalar fields, which has the baryon number $-2/3$. $\f_u \AND \tl\f_u$ are $Z_2$ odd and even, respectively. 
In addition, there are for the new fields with appropriate masses. Then, by integrating out the up-type di-quarks $\phi_u \AND \tilde{\phi}_u$ {with the same bayron number of $-2/3$}, which 
transform under $Z_2$ (odd and even, respectively), we obtain
$$
M_q^{3} \sim \frac{m_{\tilde{\f}_u}^2\, m_{\f_u}^2}{\tilde y_N \tilde y_\f\, A_\phi }.
$$
We can achieve $M_{q}$ as large as $\TEV$ if $\tilde{\f}_u$ and $\f_u$ are around the TeV scale, in order to be consistent with collider searches (see \cite{Azatov:2021irb} for a more detailed analysis of a similar model for baryogenesis). The constraint from flavor-changing neutral currents (FCNC) is alleviated because $\tilde{\f}_u$ is a diquark and tree-level FCNCs are absent~\cite{Giudice:2011ak, Azatov:2021irb}.\footnote{We can also introduce two additional copies of the scalars to ensure that flavor and CP violation are minimized, in agreement with experimental constraints.} Note that the bound from early cosmology is significantly alleviated in a low-reheating scenario, as the relics of the light exotic particles $N_{\rm dark}$ and $\pi_{\rm dark}$ are diluted by substantial entropy production. In this sense, our scenario may be consistent with current bounds and could be tested in the future. Nevertheless, a more careful analysis is warranted. 

\subsection{The highest-energy cosmic rays and mirror SM}
Thus far, we have not explained the theoretical origin of assumptions {\bf (1) \AND (3)}. Here, we show that a mirror scenario—in which, in addition to the SM, there exists a copy of the SM with a nearly identical spectrum and couplings—can account for both assumptions. Such a scenario has been discussed over decades historically~\cite{Kobzarev:1966qya} (see also reviews \cite{Berezhiani:2003xm,Okun:2006eb}). 

Assuming that there is only a single baryon number symmetry, a mixing between the dark nucleon and the nucleon can be generated. Then, we obtain the hadronic picture Lagrangian together with the dark version of the well-known charged pion interaction, 
\beq
{\cal L}\supset -\D m_{\rm mix} \bar n\, n_{\rm dark} - i y_{\rm dark} \pi_{\rm dark}   \bar p_{\rm dark} \g_5 n_{\rm dark}+h.c..
\eeq
Here, $y_{\rm dark}$ is the dark pion coupling, and $N_{\rm dark}$ is identified with $p_{\rm dark}$, while $\pi_{\rm dark}$ is a pseudo Nambu--Goldstone boson with dark electromagnetic charge. This implies that condition {\bf (3)} is automatically satisfied.

By diagonalizing the mass matrix, one obtains
\beq
{\cal L}^{\rm (canonically\mbox{-}normalized)}\supset - i\, g_{\rm dark}\,\pi_{\rm dark}\,\bar{p}_{\rm dark}\,\gamma_5\, n\,+h.c.,
\eeq
with
\beq
g_{\rm dark}\sim y_{\rm dark}\,\theta_{\rm mix}, \qquad \theta_{\rm mix}\sim \frac{\Delta m_{\rm mix}}{m_{N,{\rm dark}}-m_N}\,.
\eeq
Thus, the decay
\beq
p_{\rm dark}\to \pi_{\rm dark}+n
\eeq
is induced. The two-body decay width is given by 
\begin{equation}
  \Gamma = \frac{
  {\theta_\text{mix}^2\, y_{\rm dark}^2}}{16\pi\, m_{{p},{\rm dark}}^3}\,\lambda\Bigl(m_{{p},{\rm dark}}^2,\, m_{n}^2,\, m_{\pi,{\rm dark}}^2\Bigr)\,\Bigl[\bigl(m_{{p},{\rm dark}}-m_{n}\bigr)^2-m_{\pi,{\rm dark}}^2\Bigr]\,,
\end{equation}
which differs slightly from the corresponding expression in the real scalar case. In the small-$Q$ limit, the decay rate becomes
\beq
\Gamma \sim \frac{{\theta_\text{mix}^2\, y_{\rm dark}^2}}{2\sqrt{2}\pi}\,\frac{m_{\pi,{\rm dark}}^{3/2}\, Q^{3/2}\,\sqrt{m_n}}{m_{p,{\rm dark}}^{5/2}}\,.
\eeq
Again, under assumption {\bf (1)} with $m_{N,{\rm dark}}\sim Q\sim \GEV$, the decay rate is maximized. This condition is further supported by assuming that the mirror symmetry is only softly broken.

Given that $y_{\rm dark}\sim 4\pi$, as in ordinary QCD, the same argument as around \Eq{ldecay} applies, requiring $g_{\rm dark}\gtrsim 10^{-12}$ so that $N_{\rm dark}$ decays before the majority of it reaches Earth. This in turn implies $\theta_{\rm mix}\gtrsim 10^{-13}$. Although the non-vanishing $\theta_{\rm mix}$ contributes to neutron-dark neutron oscillations~\cite{Hostert:2022ntu}, the effect is completely negligible for $Q\sim \GEV$. Even in stellar environments with temperatures $\lesssim \MEV$, this model remains unconstrained, and the mixing is too small to be probed at the intensity frontier.

\paragraph{UHECR spectra}
In this scenario, the DM may decay into dark SM particles, which subsequently cascade into dark protons in dark sector.
The lifetime of the dark neutrons can be much shorter than that of the SM neutron if the mass difference is larger than in the SM. This can occur if the dark Higgs expectation value is larger than the SM one, leading to a greater mass difference between the dark down and dark up quarks than in the SM. Given that the $Q$-value is not very large, we expect the nucleon spectrum to be similar to that of the proton discussed in \Sec{2}. This is because the dark SM has a spectrum similar to the SM, and hence the dark proton will exhibit a flux spectrum similar to that of the proton discussed earlier in \Sec{2}.

\paragraph{UV Completion} Again, let us construct an example of a UV completion to support our discussion. To generate the mixing term, we require a dimension-9 operator, such as 
\beq
\frac{udd\, u_{\rm dark}d_{\rm dark}d_{\rm dark}}{M_{\rm mix}^5},
\eeq
in the language of quarks. This leads to an effective term of order $\sim \L_{\rm QCD}^6\bar{n} n_{\rm dark}/M_{\rm mix}^5$, implying that $M_{\rm mix} \lesssim 60\GEV$ for $\theta_{\rm mix} \gtrsim 10^{-13}$. To UV complete this higher-dimensional operator, we can consider the mirror-symmetric interaction
\beq 
{\cal L}\supset -\tl y_d\, d\, d\, \tl\f_{u} - \tl y_u\, \tl \f_u\, \bar{u} \psi_L - \tl y_u\, \tl\f_{u,\rm dark}\, \bar{u}_{\rm dark}\, \psi_R - \tl y_d\, d_{\rm dark}\, d_{\rm dark}\, \tl\f_{u,\rm dark}+h.c..
\eeq
Here, $\tl \f_u$ is the same as in the previous subsection, and $\psi$ is a gauge singlet Dirac fermion. Note that the mirror symmetry exchanges the quark and scalar quark with their dark sector counterparts, and it replaces $\psi_L \AND \psi_R$, which have the baryon number of $1, -1$, respectively. The baryon number for the dark sector counter parts are also opposite to that for the SM sector.
$L$ and $R$ denotes the chirality of $\psi$. 
Then, by integrating out $\tl \f_{u}$, $\tl \f_{u,\rm dark}$, and $\psi$, we obtain the desired dimension-9 operator, with
$$
M_{\rm mix}^{5} \sim \frac{m_{\tl\f_u}^{2}\, m_{\psi}\, m_{\tl\f_u,\rm dark}^2}{|\tl{y}_d|^2 |\tl y_u|^2}.
$$
Note that $\tilde{\phi}_u$ must have a mass above the TeV scale in accordance with LHC bounds, while $\tilde{\phi}_{u,{\rm dark}}$ and $\psi$ are not so strongly constrained. For $m_{\tl\phi_u}\sim 2\,\TEV$, one requires $m_{\psi}\, m_{\tl\phi_u}^2\lesssim (20\,\GEV)^3$ to ensure a sufficiently short decay length. Moreover, a soft breaking of the mirror symmetry (i.e., $m_{\tl\phi_u,{\rm dark}}\ll m_{\tl\phi_u}$) leads to a contribution lowering dark QCD scale via renormalization-group running. This effect can be compensated by increasing the dark Higgs mass scale, which produces a larger dark vacuum expectation value. Consequently, the dark SM quark masses are slightly increased, enhancing dark QCD scale via renormalization running effect to values comparable to those in the SM. Interestingly, this also results in a larger mass difference between the dark down and up quarks, leading to a more rapid decay of the dark nucleon into the dark proton after production.

An interesting question is whether a simultaneous explanation of the KM3-230213A event is possible. 
One possibility is to modify the model with hidden baryon number violation to explain the KM3NeT signal from the conventional UHECR-induced cascades~\cite{Alves:2025xul}.

\section{Conclusions and discussion}
In this paper, we have explored inflaton DM models that may be testable with observations of the highest-energy particles including UHECRs and UHE neutrinos. The superheavy DM mass scale can be naturally explained because the DM is identified with an axion inflaton, as in the well-known natural inflation model. Although the vanilla version of natural inflation has been excluded by the existing CMB data, by introducing a temporary “dark energy” component for reheating/second inflation, this scenario can be revived with a stable inflaton and a consistent DM abundance.

A slight violation of the dark charge conjugation symmetry, which stabilizes the inflaton DM, induces rare decays of the superheavy DM, thereby producing the highest-energy astroparticles. Since the inflaton is a heavy spin-zero particle, it is important to study both the three-body decays and the two-body decays into gauge boson pairs due to chirality arguments. These are specific features of the model that have not been thoroughly studied in the context of multi-messenger observations.
With the careful examination, we found that some of the highest-energy cosmic rays, such as the AMATERASU event, could be only marginally explained due to the stringent bounds from $\gamma$-ray observations if the DM only decays into SM particles. The KM3-230213A event itself suffers from the strong tension with the IceCube upper limit, and the combined flux point could be explained if the leptonic decay is considered. 

To avoid constraints from $\gamma$-ray observations, in addition, we constructed particle models, in which inflaton DM decays predominantly into dark particles at the $\GEV$ scale, which may subsequently decay into SM neutrons. We show that a mirror model can realize this scenario rather naturally. The prediction of this model is that the resulting highest-energy cosmic rays such as the AMATERASU event are predominantly (anti)neutrons, and thus do not suffer from deflections by galactic magnetic fields of the host galaxy. 
Therefore, a stronger correlation between the arrival directions of the highest-energy cosmic rays and locally dense regions, such as dSphs with suppressed baryonic content, would be a smoking-gun signal for this scenario. More detailed information from the highest-energy cosmic ray events such as the one around Leo I, as observed by the TA, would be useful.
Such an exotic explanation leads to a tensor-to-scalar ratio of $r>\O(0.001)$, and a running of the spectral index of $\a_s>10^{-3}$, which can be measured in future experiments. In the UV completion, we predict light new colored particles, which could be tested at collider experiments such as the HL-LHC and future colliders.

\section*{Acknowledgement}
This work is supported by JSPS KAKENHI Grant Nos. 20H05851 (W.Y.), 20H05852 (K.M.) 22K14029 (W.Y.), 22H01215 (W.Y.), Graduate Program on Physics for the Universe (Y.N.), and JST SPRING, Grant Number JPMJSP2114 (Y.N.). W.Y. is also supported by Incentive Research Fund for Young Researchers and Selective Research Fund for Young Researchers
from Tokyo Metropolitan University.
The work is also supported by NSF Grants Nos.~AST-2108466 (K.M.), AST-2108467 (K.M.), and AST-2308021 (K.M.).

\appendix 
\section{Details of calculations for three-body decay}
\label{app:1}
Since we are interested in the superheavy DM decay around the present epoch in the broken phase of the electroweak symmetry. 
Still, we will use the symmetric phase Lagrangian for the calculation because, according to the equivalence theorem, the estimation in the symmetric phase also represents the reaction of $\f$ decays into longitudinal modes of gauge bosons.

Now let us consider the $\phi$ decays into $H \bar u Q$ in the symmetric phase for illustrative purpose. The spin summed amplitude is given by 
\begin{equation}
|{\cal M}|^2 \propto\frac{1}{|M|^2} p_Q \cdot p_{u}
\end{equation}
$p_Q$ and $p_u$ are the momenta of quark $Q$ and anti-quark $\bar{u}$, respectively.

The differential decay rate for the 3-body decay channel (\ref{eq:3body}) is 
\begin{equation}
d \G_{{\f \to H \bar u Q}} = \sum_{I} \frac{1}{2 m_\f} \frac{d^3 p_H}{(2 \pi)^3 2E_H} \frac{d^3 p_Q}{(2 \pi)^3 2E_Q} \frac{d^3 p_u}{(2 \pi)^3 2E_u} |{\cal M}|^2 (2 \pi)^2 \d^{(4)}(p_\f - p_H - p_Q - p_u)
\end{equation}
Here the summation is for the $SU(2)_L$ index.

The probability of each particle with momentum $p$ to be produced can be estimated as
\begin{align}
\Phi_{H} &\equiv\frac{1}{\G_{{\f \to H \bar u Q}}}\frac{d \G_{{\f \to H \bar u Q}}}{d x_H} = 6 x (1 - x), \\
\Phi_{Q,\bar u} &\equiv\frac{1}{\G_{{\f \to H \bar u Q}}}\frac{d \G_{{\f \to H \bar u Q}}}{d x_{Q,u}} = 3 x^2,
\end{align}
where $x \equiv {2 p}/{m_\f}$, with $p$ being the total momentum of the particle of focus. The range is restricted in $x=0-1$ from kinematics. 
Now let us use the equivalent theorem to the momentum distribution of $h$, and the longitudinal modes of $W$, and $Z$, which corresponds to the NG modes in the Higgs multiples.

We are ready to discuss realistic particle spectrum for each $\phi$ decay. We note that the decay happens in the broken phase, and that $\phi$ decays also into $H^* \bar Q u$. 
In the broken phase, we have six types of the decay channels (in helicity eigenstates of quarks, for simplicity) $\f\to h \bar u_L u_R$, $\f\to W_- \bar d_L u_R, \f \to Z \bar u_L u_R$, $\f\to h u_L \bar  u_R$, $\f\to W_+  d_L \bar  u_R, \f \to Z  u_L \bar u_R$ including the decays into antiparticles, with the weight of probability of $(1,2,1,1,2,1)\times \frac{1}{8}\Gamma_{\rm 3-body}$, according to the equivalence theorem, respectively. Here $W$ and $Z$ correspond to the longitudinal modes of the gauge bosons. 

Then we get the momentum distribution, $f_i$, of particle $i$ for each $\f$ decays in the broken phase as 
\begin{align}
f_h\simeq f_{Z}&\simeq \frac{1}{4}\F_H= \frac{3}{2} x(1-x), \\ 
f_{W_+}=f_{W_-}&\simeq  \frac{1}{4}\F_H= \frac{3}{2}x(1-x),\\ 
f_{{u}_R}&\simeq f_{\bar{u}_R}\simeq \frac{1}{2}\Phi_{Q,u}= \frac{3}{2} x^2,\\ 
f_{u_L}&\simeq f_{\bar{u}_L}\simeq f_{d_L}\simeq f_{\bar{d}_L}\simeq\frac{1}{4}\F_{Q,u}=\frac{3 }{4}x^2.
\end{align}

    \bibliographystyle{apsrev4-1}
    \bibliography{ref.bib}

\end{document}